\documentclass[10pt, conference, letterpaper, table]{IEEEtran}

\IEEEoverridecommandlockouts
\usepackage{graphicx}

\ifCLASSOPTIONcompsoc
    \usepackage[caption=false, font=normalsize, labelfont=sf, textfont=sf]{subfig}
\else
\usepackage[caption=false, font=footnotesize]{subfig}
\fi
\usepackage{cite}
\usepackage{url}
\usepackage{mdwmath}  
\usepackage{soul}
\usepackage[table]{xcolor}
\usepackage{amsfonts,amsmath,amssymb,amsxtra,amsbsy,floatflt}
\usepackage{indentfirst} 
\usepackage{tabularx}
\usepackage{booktabs}
\usepackage{multirow}
\usepackage{nth}
\usepackage{comment}
\usepackage{epstopdf}
\usepackage{hhline}
\usepackage{siunitx}
\usepackage{booktabs}
\usepackage[bottom]{footmisc}
\usepackage[acronym]{glossaries}
\usepackage{upgreek}
\usepackage{algorithm}
\usepackage{algorithmicx,algpseudocode}
\usepackage{bm}

\newtheorem{problem}{Problem}
\newcommand{\vv}[1]{\mathbf{#1}}

\newcommand{\tr}{\mathrm{tr}}

\newcommand{\tran}{\mathrm{T}}
\newcommand{\herm}{\mathrm{H}}
\newcommand{\Exp}{\mathbb{E}}

\newcommand{\ssub}[1]{{\scriptscriptstyle { #1}}}

\newcommand{\name}{MARISA}

\newcommand\blfootnote[1]{%
  \begingroup
  \renewcommand\thefootnote{}\footnote{#1}%
  \addtocounter{footnote}{-1}%
  \endgroup
}

\usepackage{tikz}
\usepackage{textcomp}
\usepackage{hyperref}
\usepackage{lipsum}
\newcommand\copyrighttext{%
  \footnotesize \textcopyright 2022 IEEE. Personal use of this material is permitted.
  Permission from IEEE must be obtained for all other uses, in any current or future 
  media, including reprinting/republishing this material for advertising or promotional 
  purposes, creating new collective works, for resale or redistribution to servers or 
  lists, or reuse of any copyrighted component of this work in other works. 
  }
\newcommand\copyrightnotice{%
\begin{tikzpicture}[remember picture,overlay]
\node[anchor=south,yshift=1pt] at (current page.south) {\fbox{\parbox{\dimexpr\textwidth-\fboxsep-\fboxrule\relax}{\copyrighttext}}};
\end{tikzpicture}%
}

\newacronym{3d}{3D}{three dimensional}
\newacronym{aoa}{AoA}{angle of arrival}
\newacronym{aod}{AoD}{angle of departure}
\newacronym{ap}{AP}{access point}
\newacronym{b5g}{B5G}{Beyond-5G}
\newacronym[plural=BSs, firstplural=base stations (BSs)]{bs}{BS}{base station}
\newacronym{csi}{CSI}{channel state information}
\newacronym{dc}{DC}{direct current}
\newacronym{doa}{DoA}{direction-of-arrival}
\newacronym{emf}{EMF}{electromagnetic field}
\newacronym{em}{EM}{electromagnetic}
\newacronym{fp}{FP}{fractional program}
\newacronym[plural=HRISs, firstplural=Hybrid Reconfigurable Intelligent Surfaces (HRISs)]{hris}{HRIS}{hybrid reconfigurable intelligent surface}
\newacronym{ios}{IoS}{Internet-of-Surfaces}
\newacronym{iot}{IoT}{Internet-of-Things}
\newacronym[plural=KPIs, firstplural=key performance indicators (KPIs)]{kpi}{KPI}{key performance indicator}
\newacronym{lf}{LF}{low frequency}
\newacronym{los}{LoS}{line-of-sight}
\newacronym{mimo}{MIMO}{multiple-input multiple-output}
\newacronym{miso}{MISO}{multiple-input single-output}
\newacronym{ml}{ML}{machine learning}
\newacronym{mmse}{MMSE}{minimum mean squared error}
\newacronym{mrt}{MRT}{maximum-ratio transmission}
\newacronym{mse}{MSE}{mean squared error}
\newacronym{nlos}{NLoS}{non-line-of-sight}
\newacronym{pdf}{pdf}{probability distribution function}
\newacronym{pla}{PLA}{planar linear array}
\newacronym{pap}{P\&P}{plug-and-play}
\newacronym{ppp}{PPP}{Poisson point process}
\newacronym[plural=RISs, firstplural=Reconfigurable Intelligent Surfaces (RISs)]{ris}{RIS}{Reconfigurable Intelligent Surface}
\newacronym{rf}{RF}{radio frequency}
\newacronym{rmse}{RMSE}{root-mean-square error}
\newacronym{rss}{RSS}{received signal strength}
\newacronym{sdp}{SDP}{semidefinite programming}
\newacronym{sdr}{SDR}{semidefinite relaxation}
\newacronym{sinr}{SINR}{signal-to-interference-plus-noise ratio}
\newacronym{smse}{SMSE}{sum mean squared error}
\newacronym{snr}{SNR}{signal-to-noise ratio}
\newacronym{soa}{SoA}{state-of-the-art}
\newacronym{sre}{SRE}{smart radio environment}
\newacronym{toa}{ToA}{time-of-arrival}
\newacronym[plural=UEs, firstplural=user equipments (UEs)]{ue}{UE}{user equipment}
\newacronym{ula}{ULA}{uniform linear array}

\title{\name{}: A Self-configuring Metasurfaces Absorption and Reflection Solution Towards 6G}

\author{
	\IEEEauthorblockN{Antonio~Albanese\IEEEauthorrefmark{1}\IEEEauthorrefmark{2}, Francesco~Devoti\IEEEauthorrefmark{1}, Vincenzo~Sciancalepore\IEEEauthorrefmark{1}, Marco Di Renzo\IEEEauthorrefmark{3},  Xavier~Costa-Pérez\IEEEauthorrefmark{4}\IEEEauthorrefmark{1}}
	\IEEEauthorblockA{
		\IEEEauthorrefmark{1}NEC Laboratories Europe, Heidelberg, Germany\\
		\IEEEauthorrefmark{2}Universidad Carlos III de Madrid, Leganés, Madrid, Spain\\
		\IEEEauthorrefmark{3}Universit\'e Paris-Saclay, CNRS, CentraleSup\'elec, Laboratoire des Signaux et Syst\`emes, Gif-sur-Yvette, France\\
		\IEEEauthorrefmark{4}i2CAT Foundation and ICREA, Barcelona, Spain}
}

\begin{document}

\maketitle
\copyrightnotice

\begin{abstract}
\glspl{ris} are considered one of the key disruptive technologies towards future 6G networks. \glspl{ris} revolutionize the traditional wireless communication paradigm by controlling the wave propagation properties of the impinging signals at will.
A major roadblock for \gls{ris} is though the need for a fast and complex control channel to continuously adapt to the ever-changing wireless channel conditions. In this paper, we ask ourselves the question: \emph{Would it be feasible to remove the need for control channels for RISs?} We analyze the feasibility of devising \emph{Self-Configuring Smart Surfaces} that can be easily and seamlessly installed throughout the environment, following the new \gls{ios} paradigm, without requiring modifications of the deployed mobile network. To this aim we design \name{}, a \emph{self-configuring} metasurfaces absorption and reflection solution. Our results show that \name{} achieves outstanding performance, rivaling with state-of-the-art control channel-driven \glspl{ris} solutions.

\end{abstract}

\begin{IEEEkeywords}
5G, 6G, RIS, Self-configuring, HRIS, IRS, IoS
\end{IEEEkeywords}

\glsresetall
\section{Introduction}
\label{sec:intro}

\blfootnote{This work has been supported by EU H2020 RISE-6G project under grant agreement no. 101017011.}
The impelling need for unprecedented network performance has indicted the classical communication paradigm for not being able to fully control the propagation environment towards astounding wireless transmission efficiency targets. This has called for a revolutionary technology that could provide the means for continuously monitoring the surrounding environment and selfishly control how signal waves propagate~\cite{TAOM20}: Metasurfaces with their reflectarray-based variant, namely \gls{ris}, epitomize a fully controllable and flexible smart propagation environment built on top of man-made surfaces that alter the radio propagation properties of the impinging signals in favor of specific directions. While this game-changing technology introduces a bulk of new business opportunities and advanced use-cases for the next generation of wireless networks (B5G or 6G), it involves technical challenges that are not easy to address\cite{RZDAC_jsac20}.

The \gls{ris} paradigm turns weaknesses into strengths: the imagery of a black-box environment is transformed into a smart place packed with a massive number of surfaces that are equipped with low-cost and low-complexity electronics~\cite{RIScommag_2021}. However, an ad-hoc control channel is essential to dynamically reconfigure the propagation of radio waves based on current network dynamics. Additionally, due to the absence of signal processing units on such nearly-passive surfaces, channel estimation and acquisition are usually performed over the entire transmitter-\gls{ris}-receiver path, which may prevent the agile deployment of the surfaces\cite{di2019smart}. 

The expected high market penetration of such devices cannot impose, however, stringent requirements on their agile deployment and configuration, as these are essential features for the openness of radio access networks\cite{WhitePaperORAN}, which would put into question how easily the \gls{ris} technology will be seamlessly integrated into existing networks and standards~\cite{albanese2021rennes}. Therefore, in this paper, we take the \gls{iot} to the next level, by coining and introducing the \gls{ios} paradigm: a multitude of simple reconfigurable devices that can be seamlessly \emph{plugged} into the existing network infrastructure without requiring sophisticated installation procedures and that can autonomously \emph{play} to enhance communications \glspl{kpi}.

\begin{figure}[t]
        \center
        \includegraphics[width=\linewidth, trim = {0cm, 0.5cm, 0cm, 0cm}]{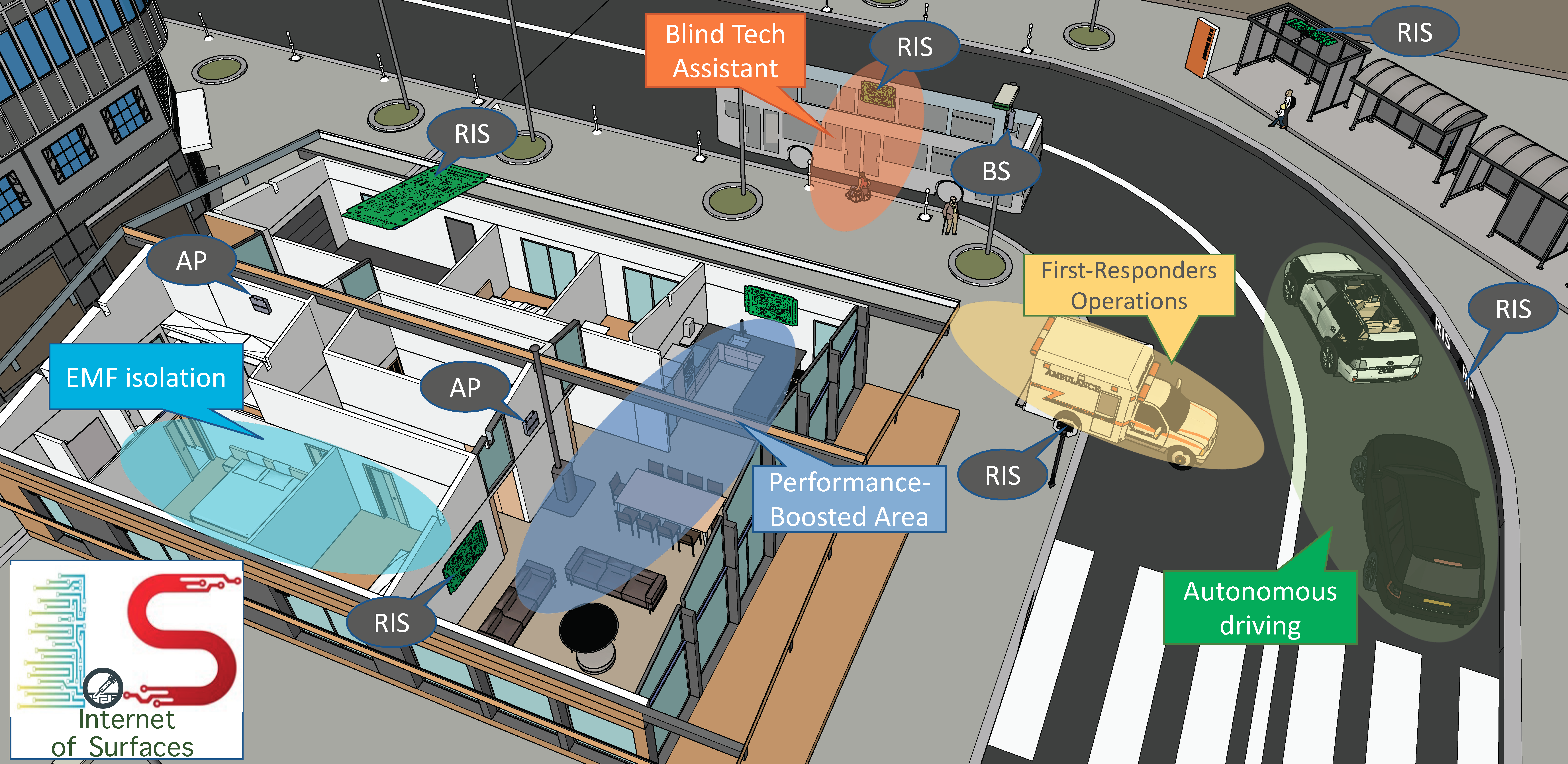}
        \caption{\acrfull{ios} reference scenario: Plug\&Play devices (\gls{ris}) to assist outdoor Base Stations (BS) and indoor Access Points (APs) in 6G envisioned use-cases.}
        \label{fig:intro_scenario}
\end{figure}

In Fig.~\ref{fig:intro_scenario}, the \gls{ios} reference scenario is depicted, wherein \glspl{ris} are densely deployed and strategically placed to extend existing customer services or to enable new ones. Indoor applications may include areas where coverage and connectivity are boosted (performance-boosted areas), as well as specific locations where the \gls{emf} exposure can be effectively reduced (\gls{emf} isolation)~\cite{BOL20_ComMag}. Interestingly, autonomous driving, e-accessibility, and first-responders operations may be supported by \gls{ris}-empowered networks without requiring the installation and maintenance of additional \glspl{ap} or \glspl{bs}~\cite{albanese2021responders}.   

Building upon the envisioned \gls{ios} paradigm, this paper presents a novel solution, namely Metasurface Absorption and Reflection for Intelligent Surface Applications (\name{}), which leverages \glspl{hris},  described in Section~\ref{sec:hris}, as plug-and-play devices with reflection and power-sensing capabilities. \name{} is built upon $i$) a new channel estimation model lato-sensu at the \gls{ris}, which is introduced in Section~\ref{sec:model_design} and $ii$) an autonomous \gls{ris} configuration methodology based only on the \gls{csi} of the \gls{bs}-\gls{ris} and \gls{ue}-\gls{ris} paths without involving an active control channel, which is described in Section~\ref{sec:absorption_model}. Finally, in Section~\ref{sec:results}, \name{} is shown to provide near-optimal performance when compared to the full-\gls{csi}-aware approach.



\emph{Notation}. We denote matrices and vectors in bold text while each of their element is indicated in roman with a subscript. $(\cdot)^{\tran}$ and $(\cdot)^{\herm}$ stand for vector or matrix transposition and Hermitian transposition, respectively. The L$2$-norm of a vector and the \emph{Frobenius} norm of a matrix are denoted as $\| \cdot \|$ and $\| \cdot \|_F$, respectively, and $\tr(\cdot)$ indicates the trace of a square matrix. Also, $\langle\cdot, \cdot \rangle$ denotes the inner product between vectors, and $\circ$ denotes the \emph{Hadamard} product between two matrices.

\section{Plug \& Play HRIS: Key Characteristics}
\label{sec:hris}

In this section, we introduce the key concept of \gls{hris} and provide a brief overview of the proposed bespoke hardware design. Then, we analyze the challenges that need to be overcome to enable self-configuration capabilities for the \gls{ios}.

\begin{figure}[t]
        \center
        \includegraphics[width=\linewidth]{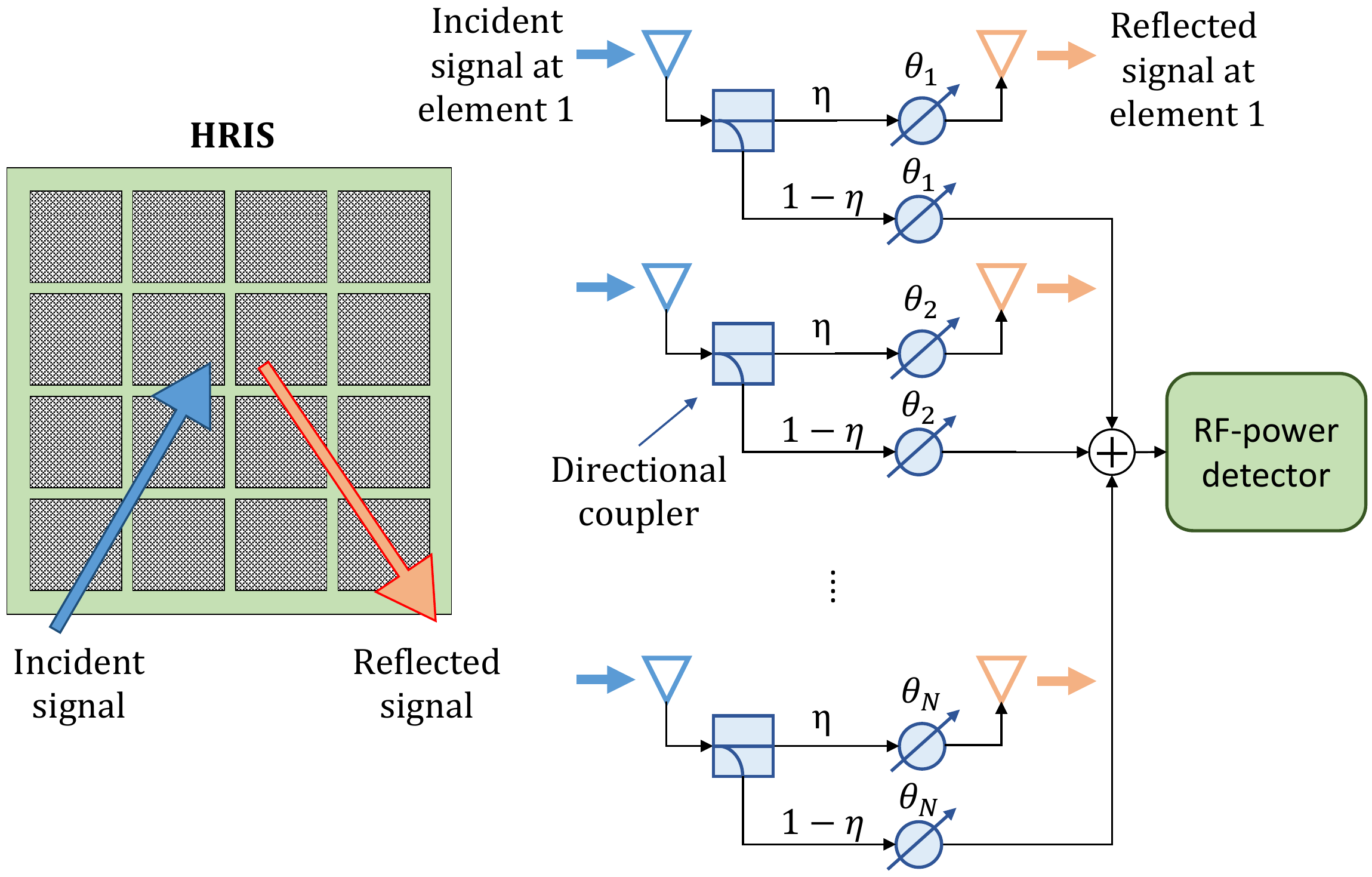}
        \caption{Reference diagram of a hybrid reconfigurable intelligent surface.}
        \label{fig:hris}
\end{figure}

\subsection{Preliminaries}
\label{sec:preliminaries}

We consider an \gls{hris}~\cite{alexandropoulos2021hybrid} comprising an array of hybrid meta-atoms, which are able to simultaneously reflect and absorb (i.e., sense the power of) incident signals. In the considered architecture, each metasurface element is coupled with a sampling waveguide that propagates the absorbed (i.e., sensed) power of the incident \gls{em} waves towards some downstream \gls{rf} hardware for enabling signal processing. 

To reduce the complexity and cost of the required hardware, the proposed \gls{pap} \glspl{hris} are not equipped with fully-fledged \gls{rf} chains but only with an \gls{rf} power detector, thereby eliminating the need for a receiver. As shown in Fig.~\ref{fig:hris}, the signals sensed by each metasurface element are summed together by some \gls{rf} combiners, which may be easily implemented as lumped components throughout the metasurface \gls{rf} circuit~\cite{lamminen200860}. The resulting signal is fed into an \gls{rf} power detector that converts the \gls{rf} power into a measurable \gls{dc} or a \gls{lf} signal, which is, e.g., made of a thermistor or a diode detector, \cite{li2020novel,yasir2019integration}. In the considered hardware architecture, the reflected and absorbed signals are subject to the same phase shifts applied by the metasurface elements, which are tuned to simultaneously control the signal reflection and power absorption properties of the \gls{hris}. Nonetheless, it is possible to enable the fully-absorption operating mode by deactivating the reflection of signals by means of simple varactor diodes~\cite{bhattacharya20198}.

\subsection{The Road Towards the IoS: Self-Configuring RISs}
\label{sec:overview}


\textbf{Managed \gls{ris} deployment.} Conventional \gls{soa} \gls{ris} deployments rely on a control channel between the \glspl{ris} and a centralized controller\footnote{Current \glspl{ris} are conventionally controlled by a centralized entity (i.e., an orchestration layer) or by the \gls{bs} itself. Within the O-RAN-compliant architecture, this might be achieved through an ad-hoc interface that is placed between the Near-Real Time (RT) RAN Intelligent Controller (RIC) and the \gls{ris} controller, which, however, leads to additional overhead.}, which serves a twofold purpose: $i$) sharing the \gls{csi} estimated at the \gls{bs} and the \glspl{ris}, $ii$) enabling the joint optimization of the \gls{bs} precoding matrix and the phase shifts at the \gls{ris} elements, in order to avoid losses due to the out-of-phase reception of uplink signals at the \gls{bs} or downlink signals at the \acrfullpl{ue}. Indeed, if both the direct and reflected (through one or multiple \glspl{ris}) propagation links between the \gls{bs} and a \gls{ue} are available, the transmission delays experienced by the transmit signals over the two paths may be substantially different, thereby requiring the \gls{ris} to be configured for compensating them. Such configuration is feasible only if the centralized control entity has perfect \gls{csi} of the direct and reflected propagation channels, as well as it has full control on the \gls{ris} configuration. 

\textbf{Unmanaged \gls{hris} deployment.} Avoiding the need for an external management and control entity has major positive implications for the design and deployment of \gls{ris}-aided wireless networks. In the \gls{ios} landscape showcased in Fig.~\ref{fig:intro_scenario}, in fact, we envision that novel devices like the \glspl{hris} will be completely autonomous and self-configuring without requiring an external control channel, thereby maximizing the agility and flexibility of their deployment while keeping the installation, configuration and maintenance costs affordable. As currently available implementations of \glspl{ris} cannot operate without an external control channel, we leverage on the concept of \gls{hris} to propose the \textit{first-of-its-kind} solution that does not rely on the existence of a remote control channel but is built upon the optimization and configuration of the \gls{hris} uniquely based on local estimates of the \gls{csi} at the \gls{hris} itself. 

\textbf{Power-based indirect beamforming.} 
As shown in Fig.~\ref{fig:hris}, the proposed \gls{hris} design includes only one \gls{rf} power detector, which can only measure the power of all the incident signals at every \gls{hris} meta-atom. Most available \gls{aoa} estimation techniques necessitate the signal samples at each receive antenna. Conversely, we make the most out of our proposed hardware design and perform an indirect estimation of the \gls{aoa}, by optimizing the phase shifts applied to the absorbed (sensed) signals at every meta-atom so as to maximize the power sensed by the detector. As adjusting the phase shifts applied by the meta-atoms is equivalent to realizing a virtual (passive) beamformer towards specified \gls{aoa} of the incident signals with respect to the \gls{hris} surface, we can take advantage of the power sensing capability of the \glspl{hris} for estimating the \gls{bs}-\gls{hris} and \gls{hris}-\gls{ue} channels with very little local information. 
 
\textbf{\gls{hris} self-configuration.} The optimal configuration of the \glspl{hris} is discussed in Section~\ref{sec:hris_config}. Here, we anticipate that any algorithmic solutions for optimizing the \gls{hris} require the estimation of the \gls{csi} of the \gls{bs}-\gls{hris} and \gls{hris}-\gls{ue} channels. This results in a chicken-egg problem that needs to be tackled. To this end, we devise an online optimization approach that relies upon a finite set of \gls{hris} configurations, namely a codebook, that can be iteratively tested for probing a finite set of predefined \glspl{aoa}. It is worth mentioning that this operation may have a disruptive impact on the network operation: a given \gls{hris} configuration may be in use for assisting, through smart reflections, the data transmission of some \glspl{bs} and \glspl{ue}. Therefore, changing the \gls{hris} configuration for sensing may negatively affect the communication performance. \name{} addresses this issue by means of a simultaneous hybrid probing and communication scheme, as detailed in Section~\ref{sec:algorithm}.

\section{Model Design}
\label{sec:model_design}
In this section, we introduce the reference analytical model for the considered wireless \gls{hris}-aided network scenario. We start by stating the \gls{hris} configuration problem accounting for the practical limitations of such hardware platform. 
\subsection{System Model}
\label{sec:sys_model}

 \begin{figure}[t]
        \center
        \includegraphics[width=.81\linewidth]{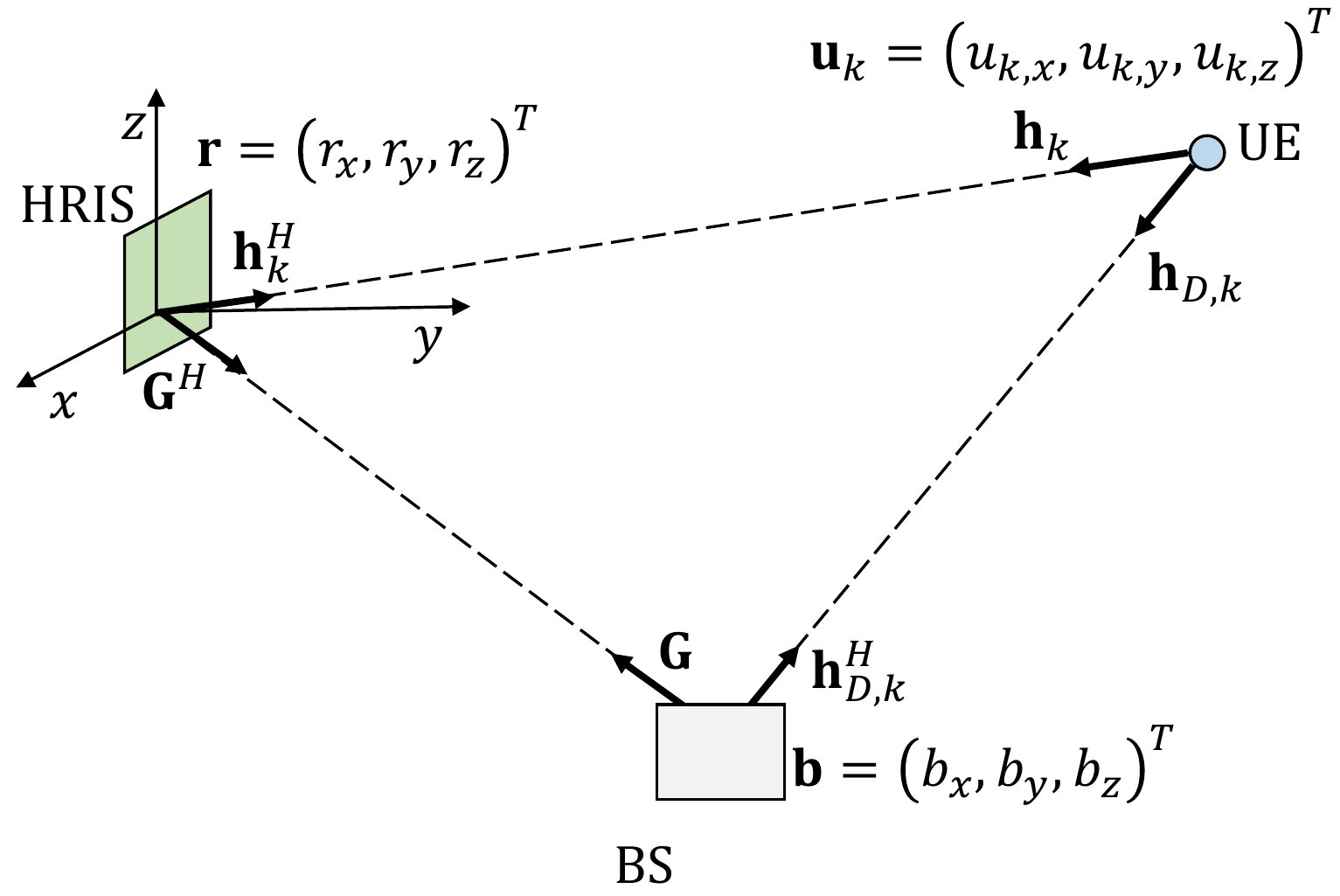}
        \caption{\label{fig:geometry} Geometrical representation of the considered scenario including the \gls{bs}, the \gls{ris} and the \gls{ue}.}
\end{figure}

We consider the scenario depicted in Fig.~\ref{fig:geometry}, in which a \gls{bs} equipped with $M$ antennas serves $K$ single-antenna \glspl{ue} with the aid of an \gls{hris}. We model the \gls{bs} as a \gls{ula}, and the \gls{hris} as a \gls{pla} equipped with $N = N_x \times N_z$ meta-atoms, where $N_x$ and $N_z$ denote the number of elements along the $x$ and $z$ axis, respectively. We assume that the inter-distance of the \gls{bs} and \gls{hris} array elements is $\lambda/2$, where $\lambda = c/f_c$ denotes the carrier wavelength, $f_c$ is the corresponding carrier frequency and $c$ is the speed of light. The joint reflection and absorption capabilities of the \gls{hris} are realized through directional couplers whose operation is determined by the parameter $\eta\in[0,1]$, which is the fraction of the received power that is reflected for communication, while $1-\eta$ is the amount of absorbed power. A practical implementation of this architecture can be found in \cite{alexandropoulos2021hybrid}.

We denote by $\vv{b}\in \mathbb{R}^3$, $\vv{r}\in \mathbb{R}^3$ and $\vv{u}_k \in \mathbb{R}^3$ the locations of the \gls{bs} center, the \gls{hris} center and the $k$-th \gls{ue}, respectively. Focusing on the downlink, the \gls{bs} transmits data to the $k$-th \gls{ue} over a direct \gls{los} link $\vv{h}_{\ssub{D},k} \in \mathbb{C}^{M\times 1}$ and a reflected link through the \gls{hris}. Such path can be decomposed into the \gls{los} channel $\vv{h}_k \in \mathbb{C}^{N \times 1}$ through which the \gls{hris} reflects the impinging signal towards the \gls{ue}, and the \gls{los} channel $\vv{G} \in \mathbb{C}^{N\times M}$ between the \gls{bs} and the \gls{hris}. The array response vector at the \gls{bs} towards the location $\vv{p} \in \mathbb{R}^3$ is denoted by $\vv{a}_{\ssub{BS}}(\vv{p}) \in \mathbb{C}^{M \times 1}$ whose elements are defined as 
\begin{equation}
    \{\vv{a}_{\ssub{BS}}(\vv{p})\}_{m=1}^M \triangleq e^{j\langle\vv{k}_{\ssub{PB}}, (\vv{b}_m - \vv{b})\rangle},
\end{equation}
where $\vv{k}_{\ssub{BP}}$ is the wave vector, defined as
\vspace{-1mm}\begin{equation}
    \vv{k}_{\ssub{PB}} \triangleq \frac{2\pi}{\lambda}\frac{\vv{p}-\vv{b}}{\|\vv{b} - \vv{p}\|},
\end{equation}
with $\vv{b}_m$ denoting the coordinates of the $m$-th \gls{bs} antenna element. 

Likewise, the \gls{hris} array response vector towards the location $\vv{p}$ is denoted by $\vv{a}_{\ssub{R}}(\vv{p}) \in \mathbb{C}^{N \times 1}$ whose elements are
\begin{equation}
    \{\vv{a}_{\ssub{R}}(\vv{p})\}_{n=1}^N \triangleq e^{j\langle\vv{k}_{\ssub{PR}} , (\vv{r}_n - \vv{r})\rangle},
\end{equation}
where $\vv{k}_{\ssub{PR}}$ is the corresponding wave vector 
\begin{equation}
    \vv{k}_{\ssub{PR}} \triangleq \frac{2\pi}{\lambda}\frac{\vv{p}-\vv{r}}{\|\vv{r} - \vv{p}\|},
\end{equation}
and $\vv{r}_n$ is the coordinate of the $n$-th meta-atom of the \gls{hris}. 

The overall gain of a generic communication path between two given locations $\vv{p}$, $\vv{q} \in \mathbb{R}^3$ is defined as
\vspace{-1mm}\begin{align}
\gamma(\vv{p},\vv{q}) \triangleq \gamma_0 \left( \frac{d_0}{\|\vv{p} - \vv{q}\|} \right)^\beta,
\label{eq:path_gain}
\end{align}
where $\gamma_0$ is the channel power gain at a reference distance $d_0$ and $\beta$ is the pathloss exponent. Hence, we can write the \gls{bs}-\gls{hris} and the \gls{hris}-$\text{\gls{ue}}_k$ channels as
\begin{align}
    \vv{G} &\triangleq \sqrt{\gamma(\vv{b},\vv{r})}\vv{a}_{\ssub{R}}(\vv{b})\vv{a}^{\herm}_{\ssub{BS}}(\vv{r}) \in \mathbb{C}^{N\times M}, \label{eq:channel_bs_ris}\\
    \vv{h}_k &\triangleq \sqrt{\gamma(\vv{u}_k,\vv{r})}\vv{a}_{\ssub{R}}(\vv{u}_k) \in \mathbb{C}^{N\times 1}, \label{eq:channel_ris_ue}
\end{align}
while the direct \gls{bs}-$\text{\gls{ue}}_k$ channel is
\begin{equation}
    \vv{h}_{\ssub{D},k} \triangleq \sqrt{\gamma(\vv{b},\vv{u}_k)}\vv{a}_{\ssub{BS}}(\vv{u}_k) \in \mathbb{C}^{M\times 1}.
    \label{eq:channel_bs_ue}
\end{equation}

Thus, the received signal at the $k$-th \gls{ue} is
\begin{equation}
y_k=\left(\sqrt{\eta}\vv{h}_k^{\herm}\vv{\Theta}\vv{G} +\vv{h}^{\herm}_{\ssub{D},k}\right)\vv{W} \vv{s} + n_k \in \mathbb{C},
\end{equation}
where $\vv{\Theta} = \mathrm{diag}[\alpha_1 e^{j\theta_1}, \dots, \alpha_N e^{j\theta_N}]$, with $\theta_i \in [0, 2\pi]$ and $|\alpha_i|^2 \leq 1$, $\forall i$ being the phase shifts and the gains introduced by the \gls{hris}, $\vv{W} \in \mathbb{C}^{M \times K}$ is the transmit precoding matrix whose $k$-th column $\vv{w}_k$ is the transmit precoder of $\text{\gls{ue}}_k$, $\vv{s} = [s_1, \dots, s_K]^\mathrm{T}$ is the transmit symbol vector with $\mathbb{E}[|s_k|^2] = 1$ $\forall k$, and $n_k$ is the noise term whose distribution is $\mathcal{CN}(0,\sigma_n^2)$. For tractability, we assume that the same \gls{hris} configuration (i.e., the phase shifts in Fig.~\ref{fig:hris}) is applied to the incident signals to compute the absorbed and reflected power. Although the two branches of the directional couplers in Fig.~\ref{fig:hris} may have dedicated phase shifters, we assume that they are the same. This assumption allows us to find a simple and useful relationship between the optimal phase shift configuration for both the reflection and absorption functions, which is beneficial for optimizing the \gls{hris}. Moreover, we assume that $\{\alpha_i\}_{i=1}^N$ and $\{\theta_i\}_{i=1}^N$ can be independently optimized.

\subsection{\gls{hris} Optimization}
\label{sec:hris_config}

In this section, we focus our attention on how an \gls{hris} can be endowed with self-configuring capabilities, and, in particular, how the absence of a dedicated control channel results in the need for the \gls{hris} of locally estimating the channels towards the \gls{bs} and the \gls{ue}, in order to establish and maintain a high-quality reflected path. To this end, we commence by formulating the optimization problem without imposing the absence of the control channel, and we then elaborate on the difficulty of solving the obtained problem by relying only on local \gls{csi} at the \gls{ris}.


\begin{figure}[t]
        \center
        \includegraphics[width=.95\linewidth]{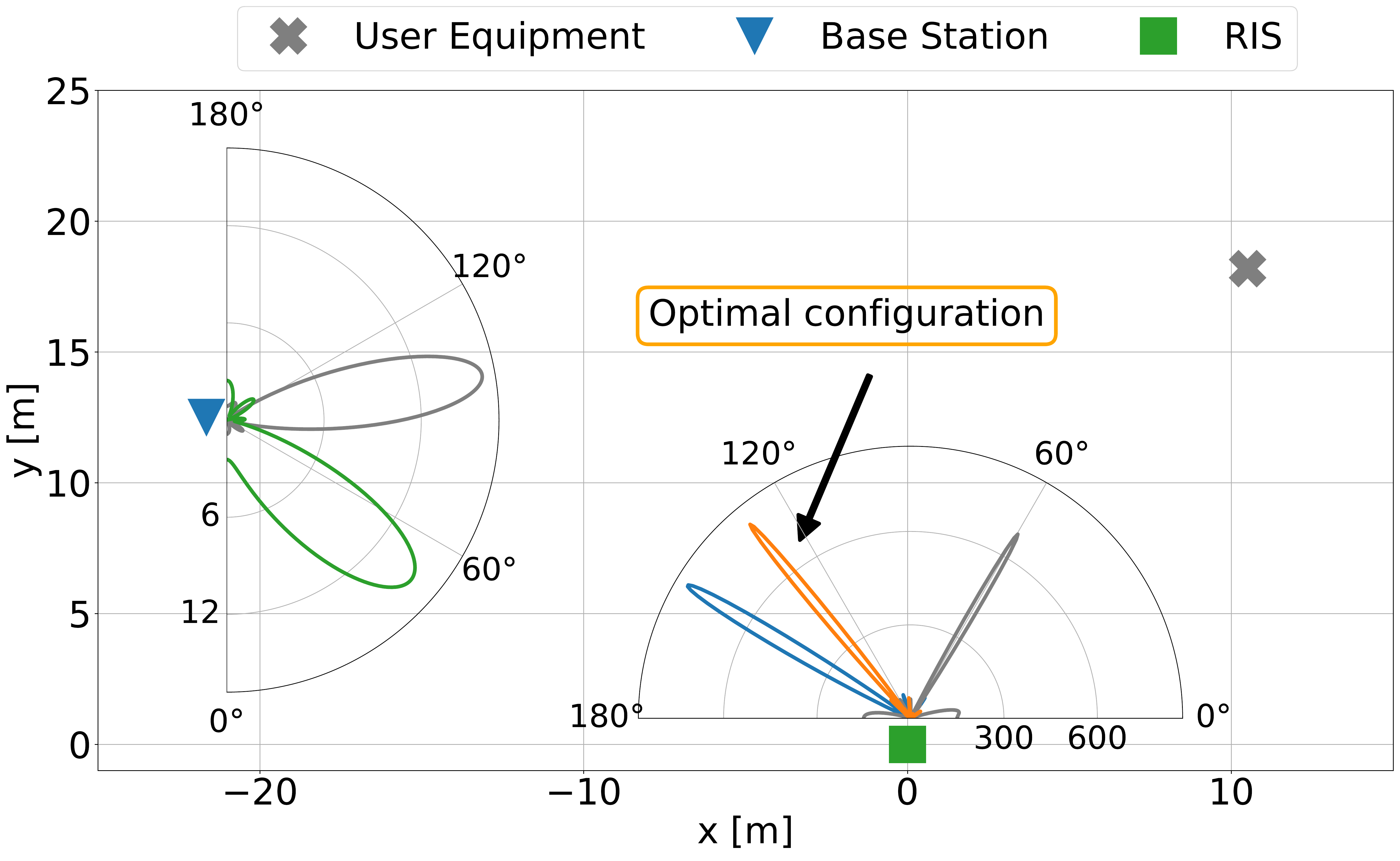}
        \vspace{-3.5mm}
        \caption{\label{fig:beam_pattern}Toy scenario with superimposed \gls{bs} and \gls{ris} antenna diagrams. The beampatterns pointing towards the \gls{bs} and \gls{ue} are respectively given by \eqref{eq:v_B} and \eqref{eq:v_U}, while the optimal \gls{ris} configuration in orange is derived as per \eqref{eq:v_BU}.}
\end{figure}

The \gls{sinr} at the $k$-th \gls{ue} can be written as 
\begin{align}
     \!\!\!\!\!\!\mathrm{SINR}_k & = \frac{\big| \big(\sqrt{\eta}\vv{h}_k^{\herm} \vv{\Theta}\vv{G} + \vv{h}_{\ssub{D}.k}^{\herm})\vv{w}_k\big|^2}{\sigma^2_n + \sum_{j \neq k}\big|\big(\sqrt{\eta} \vv{h}_k^{\herm} \vv{\Theta}\vv{G} + \vv{h}_{\ssub{D}.k}^{\herm})\vv{w}_j \big|^2}, \label{eq:sinr_k}
\end{align}
where $\vv{w}_k$ is assumed to be given during the optimization of the configuration of the \gls{hris}. More precisely, $\vv{w}_k$ is optimized by the \gls{bs} after the channel estimation phase. The disjoint optimization of $\vv{w}_k$ and the \gls{ris} configuration facilitates the design and deployment of a control channel-free \gls{hris}, which is the focus of this paper. The optimization of $\vv{w}_k$ is elaborated in further text. We aim at finding the optimal \gls{hris} configuration that maximizes the network sum-rate, which is directly related to the \gls{sinr} at every \gls{ue}, as exemplified in the toy scenario of Fig.~\ref{fig:beam_pattern}. More precisely, the network sum-rate is defined as
\begin{align}
    \mathrm{R} \hspace{-0.05cm} \triangleq \hspace{-0.1cm} \sum_{k=1}^K \log_2 \hspace{-0.1cm}\left( \hspace{-0.1cm} 1 \hspace{-0.05cm} + \hspace{-0.05cm} \frac{\big| \big(\sqrt{\eta}\vv{h}_k^{\herm} \vv{\Theta}\vv{G} + \vv{h}_{\ssub{D}.k}^{\herm})\vv{w}_k\big|^2}{\sigma^2_n \hspace{-0.05cm} + \hspace{-0.05cm}\sum\limits_{j \neq k}\big|\big(\sqrt{\eta} \vv{h}_k^{\herm} \vv{\Theta}\vv{G} \hspace{-0.05cm}+\hspace{-0.05cm} \vv{h}_{\ssub{D}.k}^{\herm})\vv{w}_j \big|^2} \hspace{-0.1cm} \right),
    \label{eq:sum_rate}
\end{align}
which results in the following optimization problem.
\begin{problem}
[\gls{sinr}-based \gls{hris} configuration]
\label{problem:max_sinr}
\begin{align}
    \max_{\vv{\Theta}} & \ \ \mathrm{R} 
    \nonumber\\
   \textup{s.t.} & \ \  |\Theta_{ii}|^2 \leq  1 \quad i = 1,\dots, N. \nonumber
\end{align}
\end{problem}

Problem~\ref{problem:max_sinr} is a \gls{fp}, since it falls into the family of optimization problems involving at least one ratio of two functions. To tackle it, we apply the \gls{fp} Quadratic Transform method~\cite{Shen2018I} to the objective function in \eqref{eq:sum_rate}, and obtain the following equivalent optimization problem.
\begin{problem}
[\gls{sinr}-based \gls{hris} configuration reformulated]
\label{problem:max_sinr_eq}
\begin{align}
    \max_{\vv{\Theta},\mu_k} & \ \ \log_2\left(2\mu_k\sqrt{A_k(\vv{\Theta})} - \mu^2_kB_k(\vv{\Theta}) \right)
    \nonumber\\
   \textup{s.t.} & \ \  |\Theta_{ii}|^2 \leq  1 \quad i = 1,\dots, N, \nonumber \\
   & \ \ \mu_k \in \mathbb{R} \quad k = 1,\dots,K, \nonumber
\end{align}
\end{problem}
where 
\begin{align}
    A_k(\vv{\Theta}) \triangleq \sum_{j = 1}^K \big| \big(\sqrt{\eta}\vv{h}_k^{\herm} \vv{\Theta}\vv{G} + \vv{h}_{\ssub{D}.k}^{\herm})\vv{w}_j\big|^2 + \sigma_n^2,
\end{align}
\begin{align}
    B_k(\vv{\Theta}) \triangleq \sum_{j \neq k} \big| \big(\sqrt{\eta}\vv{h}_k^{\herm} \vv{\Theta}\vv{G} + \vv{h}_{\ssub{D}.k}^{\herm})\vv{w}_j\big|^2 + \sigma_n^2. \label{eq:B_k}
\end{align}

Problem~\ref{problem:max_sinr_eq} has the notable property of being convex in the variables $\mu_k$ and $\vv{\Theta}$ separately. Therefore, it can be efficiently (but sub-optimally) tackled by alternating the optimization over $\vv{\Theta}$ and over the auxiliary variables $\{\mu_k\}_{k=1}^K$. However, it cannot be locally solved at the \gls{hris} for three reasons: \textit{i)} the lack of digital signal processing units at the surface, \textit{ii)} the capability of the \gls{hris} to execute only power measurements, and, more importantly, \textit{iii)} because Problem~\ref{problem:max_sinr_eq} requires the \gls{csi} of the direct link between the \gls{bs} and each \gls{ue}, which can be obtained only through a control and feedback channel.


These considerations call for an alternative approach that takes into account the constraints on the design and deployment of \glspl{hris}, while taking advantage of the massive availability of \glspl{hris} in \gls{ios}-based networks.
A solution that fulfills these requirements can be obtained by direct inspection of the \gls{sinr} in \eqref{eq:sinr_k}, and by taking into account the limitations of the \gls{hris} in terms of \gls{rf} and signal processing hardware, as well as the absence of a control channel. 

A feasible strategy to optimize the \gls{sinr} for every \gls{ue} is to optimize the \gls{hris} configuration so that the intensity of  $\vv{h}_k^{\herm}\vv{\Theta}\vv{G}\vv{w}_k$ is maximized, $\forall k$ i.e., the end-to-end \gls{ris}-assisted channel gain of each user is enhanced. This approach is approximately equivalent to maximizing the $\text{\gls{sinr}}_k$ while ignoring the interfering term $B_k(\vv{\Theta})$ in \eqref{eq:B_k}. Upon completion of this optimization, the \gls{bs} can optimize the precoding matrix $\vv{W}$, in order to co-phase the direct and the path reflected through the \gls{hris}. Indeed, even though the 
\gls{bs} cannot control the \gls{hris} configuration due to the absence of a control channel, it can always estimate the direct channel towards each \gls{ue} and the equivalent RIS-assisted link.

\textbf{Channel estimation and \gls{hris} configuration.} Let us now focus our attention on the optimization of the \gls{hris} configuration, by taking into account that it is equipped with a single \gls{rf} power detector. To this end, we  commence by deriving a closed-form expression for the \gls{hris} configuration that maximizes the reflected power.
We assume that a training phase exists, during which the \gls{bs} and each \gls{ue} transmit a pilot symbol $s$ in order to realize the initial beam alignment procedure\footnote{This standard procedure is essential before data transmission in, e.g., millimeter-wave networks for initial device discovery and channel estimation~\cite{ANR17_TWC}.}. Without loss of generality, we assume a certain degree of synchronization, i.e., the \gls{bs} and the \glspl{ue} transmit at different times, but all \glspl{ue} transmit simultaneously. We will relax this latter assumption in Section~\ref{sec:algorithm}.    

For ease of presentation, we define the vector
\begin{equation}
    \vv{v} \triangleq [\alpha_1 e^{-j\theta_1}, \dots, \alpha_N e^{-j\theta_N}]^\mathrm{T} \in \mathbb{C}^{N \times 1},
\end{equation}
such that $\vv{\Theta} = \mathrm{diag}(\vv{v}^{\herm})$. The signals at the output of the \gls{rf} combiner, which are obtained from the pilot signals transmitted by the \gls{bs} and the \glspl{ue}, can be formulated as
\begin{align}
    y_{\ssub{B}} & =   \sqrt{(1-\eta)} \, \vv{v}^{\herm} \vv{G} \vv{w}_{\ssub{R}} s  + n \in \mathbb{C}, \\
    y_{\ssub{U}} & =  \sqrt{(1-\eta)} \, \vv{v}^{\herm} \vv{h}_{\ssub{\Sigma}} s  + n \in \mathbb{C},
\end{align}
where we assume that the \gls{bs} and the \glspl{ue} emit the same amount of power $P$ and $n \sim \mathcal{N}(0,\sigma_n^2)$ is the additive noise term. 
Let $\vv{w}_{\ssub{R}}$ be the optimal \gls{bs} precoder for the \gls{bs}-\gls{hris} link. We will see shortly that the knowledge of $\vv{w}_{\ssub{R}}$ is not explicitly needed to optimize the \gls{hris} configuration. Also, we define $\vv{h}_{\ssub{\Sigma}} \triangleq \sum_{k=1}^K \vv{h}_k$. Since the \gls{ue}-\gls{hris} channel $\vv{h}_k$ corresponds to the uplink, to use it in the downlink, we assume that the channel reciprocity holds.  

Therefore, the detected power $P_{\ssub{B}}$ and $P_{\ssub{U}}$ from the pilot signals emitted by the \gls{bs} and the \glspl{ue}, respectively, can be formulated as
\begin{align}
    P_{\ssub{B}} & = \Exp \left[| y_b|^2\right]  = (1-\eta) \, \big|\vv{v}^{\herm}\vv{G} \vv{w}_{\ssub{R}} s\big|^2 + \sigma_n^2, \label{eq:p_BS}\\
    P_{\ssub{U}} & = \Exp\left[| y_u|^2\right]  =  (1-\eta) \, \big|\vv{v}^{\herm} \vv{h}_{\ssub{\Sigma}}s\big|^2 + \sigma_n^2 \label{eq:p_UE}.
\end{align}

In order to be self-configuring, an \gls{hris} needs to infer the channels $\vv{G}$ and $\vv{h}_\ssub{\Sigma}$ only based on $P_{\ssub{B}}$ and $P_{\ssub{U}}$ in \eqref{eq:p_BS} and \eqref{eq:p_UE}, respectively.
This is equivalent to finding the configuration of the \gls{hris} that maximizes $P_{\ssub{B}}$ and $P_{\ssub{U}}$, which in turn corresponds to estimating the directions of incidence of the signals on the \gls{hris}.
As a result, we formulate the following optimization problem, whose solution is the \gls{hris} configuration that maximizes $P_{\ssub{B}}$
\begin{align}
    \max_{\vv{v}} & \ \ |\vv{v}^{\herm}\vv{G}\vv{w}_R|^2 \label{eq:obj_hris_conf_BS} \\
    \textup{s.t.} & \ \ |{v}_i|^2\leq 1 \quad i = 1,\dots,N. \nonumber
    \label{eq:hris_conf_BS}
\end{align}
where ${v}_i$ is the $i$th element of $\vv{v}$.

The objective function in~\eqref{eq:obj_hris_conf_BS} can be recast as \begin{align}
    |\vv{v}^{\herm}\vv{G}\vv{w}_R|^2 = \vv{v}^{\herm}\vv{a}_{\ssub{R}}(\vv{b})\vv{a}_{\ssub{R}}^{\herm}(\vv{b})\vv{v} \, \large|z_{\ssub{R,R}}\large|^2,
\end{align}
where $z_{\ssub{R,R}} \triangleq \sqrt{\gamma(\vv{b},\vv{r})}\vv{a}_{\ssub{BS}}^{\herm}(\vv{r})\vv{w}_\ssub{R} \in \mathbb{C}$ is the projection of the \gls{bs} precoding vector $\vv{w}_{\ssub{R}}$ onto the \gls{bs}-\gls{hris} direction. 

Hence, the optimal \gls{hris} configuration for maximizing the absorbed power from the \gls{bs} is $\vv{v}_{\ssub{B}} \in \mathbb{C}^{N \times 1}$ with
\begin{align}
v_{\ssub{B},i} = e^{j \angle a_{\ssub{R},i}(\vv{b})} \quad i = 1,\dots,N. \label{eq:v_B}
\end{align}

Analogously, the optimal \gls{hris} configuration that maximizes $P_{\ssub{U}}$ is $\vv{v}_{\ssub{U}} \in \mathbb{C}^{N \times 1}$ with
\begin{align}
v_{\ssub{U},i} = e^{j \angle h_{\ssub{\Sigma},i}} \quad i = 1,\dots,N. \label{eq:v_U}
\end{align}

From \eqref{eq:v_B} and \eqref{eq:v_U}, we evince, as anticipated, that the optimal \gls{hris} configuration that maximizes the sensed power depends only on the \gls{hris} array response vectors towards the \gls{bs} and \gls{ue} directions, but it is independent of the (optimal) \gls{bs} precoding vector.

Based on $\vv{v}_{\ssub{B}}$ and $\vv{v}_{\ssub{U}}$, we are now in the position of proposing a distributed approach for optimizing the \gls{hris}. In particular, we formulate the following optimization problem.
\begin{problem}[Multi-\gls{ue} \gls{sinr}-based \gls{hris} configuration]\label{problem:max_sinr_multi_ue} 
\begin{align}
    \max_{\vv{\Theta}} & \ \ \frac{\big|\vv{h}_{\ssub{\Sigma}}^{\herm} \vv{\Theta} \vv{G}  \vv{w}\big|^2}{\sigma_n^2}    \label{eq:max_sinr_multi_ue}\\
   \textup{s.t.} & \ \  |\Theta_{ii}|^2 \leq  1 \quad i = 1,\dots, N. \nonumber 
\end{align}
\end{problem}

Problem \ref{problem:max_sinr_multi_ue} is independent of the direct channels between the \gls{bs} and the \glspl{ue}, as well as of the \gls{bs} precoder $\vv{w}$: these are fundamental requirements due to the lack of control channel. With the aid of Cauchy–Schwarz's inequality, we obtain  
\begin{equation}
    \big|\vv{h}_{\ssub{\Sigma}}^{\herm} \vv{\Theta} \vv{G}  \vv{w}\big|^2 \leq \sum_{k=1}^K \big|\vv{h}_{k}^{\herm} \vv{\Theta} \vv{G}  \vv{w}\big|^2, \label{eq:cauchy}
\end{equation}
implying that the objective function in \eqref{eq:max_sinr_multi_ue} is a lower bound for the sum of the powers of the signals transmitted by the \glspl{ue} independently, which are sensed by the \gls{hris}. Notably, the inequality in \eqref{eq:cauchy} becomes an equality if and only if the channels $\vv{h}_k$ are orthogonal to each other. 

By using \eqref{eq:cauchy}, the objective function in~\eqref{eq:max_sinr_multi_ue} can be reformulated as
\begin{equation}
    \frac{\big|z_{\ssub{R}}\vv{v}^{\herm}\hat{\vv{h}}\big|^2}{\sigma_n^2}, \label{eq:sinr_multi_ue_2}
\end{equation}
where $\hat{\vv{h}} \triangleq \vv{h}^*_{\ssub{\Sigma}} \circ \vv{a}_\ssub{R}(\vv{b})$ is the  equivalent channel that accounts for the overall effect of the aggregate \gls{ue}-\gls{hris} channels from the \gls{hris} standpoint, and $z_{\ssub{R}} \triangleq \sqrt{\gamma(\vv{b},\vv{r})}\vv{a}_{\ssub{BS}}^{\herm}(\vv{r})\vv{w} \in \mathbb{C}$ is the reflected path between the \gls{bs} and the \gls{hris} for a given precoder $\vv{w}$ at the \gls{bs}. 

Therefore, the \gls{hris} optimal configuration solution of Problem \ref{problem:max_sinr_multi_ue} is
\begin{align}
    \vv{v}_{\ssub{BU}}=e^{j\angle \  \hat{\vv{h}}}=e^{j\angle \  (\vv{h}_{\ssub{\Sigma}}^* \circ \vv{a}_{\ssub{R}}(\vv{b}))} = \vv{v}_{\ssub{U}}^* \circ \vv{v}_{\ssub{B}}, \label{eq:v_BU}
\end{align}
which proves that the \gls{hris} configuration in the absence of a control channel can be inferred solely from $\vv{v}_{\ssub{U}}$ and $\vv{v}_{\ssub{U}}$.

\textbf{Single-\gls{ue} scenario.} In the single-\gls{ue} scenario, some interesting conclusions can be drawn. In this case, the \gls{snr} can be written as
\begin{align}
    \mathrm{SNR} = \frac{|\left(\vv{h}^{\herm}\vv{\Theta}\vv{G} +\vv{h}^{\herm}_{\ssub{D}}\right)\vv{w}|^2}{\sigma_n^2},
    \label{eq:snr_single_ue}
\end{align}
where we have omitted the \gls{ue} index $k$ for simplicity. Similar to \eqref{eq:sinr_multi_ue_2}, the numerator in \eqref{eq:snr_single_ue} can be reformulated as
\begin{align}
    &\large|z_{\ssub{R}}\vv{v}^{\herm}\hat{\vv{h}} + z_{\ssub{D}}\large|^2,
    \label{eq:snr_single_ue_2}
\end{align}
with $\hat{\vv{h}}=\vv{h}^* \circ \vv{a}_{\ssub{R}}(\vv{b})$ and $z_\ssub{D} = \vv{h}^{\herm}_\ssub{D} \vv{w}$. With the aid of some algebraic manipulations, \eqref{eq:snr_single_ue_2} simplifies to
\begin{align}
    & |z_{\ssub{R}}|^2\vv{v}^{\herm}\hat{\vv{h}}\hat{\vv{h}}^{\herm}\vv{v} + |z_{\ssub{D}}|^2 + 2\mathrm{Re}\{z_{\ssub{R}}\vv{v}^{\herm}\hat{\vv{h}}z_{\ssub{D}}^*\},
    \label{eq:snr_single_ue_3}
\end{align}
which elucidates that the optimal \gls{hris} configuration needs to fulfill two conditions: $i$) the maximization of the reflected path gain $\vv{v}^{\herm}\hat{\vv{h}}\hat{\vv{h}}^{\herm}\vv{v}$ and $ii$) the phase alignment between the direct and reflected paths, i.e., $z_{\ssub{D}}$ and $z_{\ssub{R}}\vv{v}^{\herm}\hat{\vv{h}}$.

Finally, we note that $\vv{v}_{\ssub{BU}}$ in \eqref{eq:v_BU} is optimal only in the absence of the \gls{los} path, $z_{\ssub{D}}$, between the \gls{bs} and the \gls{ue}. When the \gls{bs}-\gls{ue} link is not negligible, the optimality of $\vv{v}_{\ssub{BU}}$ is guaranteed if and only if the direct and reflect paths are aligned in phase. 

\section{Codebook-Based Optimization of \glspl{hris}}
\label{sec:absorption_model}
Problem~\ref{problem:max_sinr_multi_ue} provides us with a mathematical model for optimizing the \gls{hris} configuration. From a practical standpoint, however, the optimal solution $\vv{v}_{\ssub{BU}}$ in \eqref{eq:v_BU} depends on the array response vectors from the \gls{hris} towards the \gls{bs} and the \glspl{ue}. To implement the obtained solution, the array response vectors, i.e., the \gls{bs}-\gls{ris} and \gls{ris}-\glspl{ue} \glspl{aoa}, need to be estimated, but this is not possible at the \gls{hris} due to the absence of \gls{rf} chains and of a control channel. In this section, we propose a codebook-based approach for estimating the necessary \glspl{aoa} and then computing  $\vv{v}_{\ssub{BU}}$ in a distributed manner and locally at the \gls{hris}, i.e., our proposed \name{}.

\subsection{\name{}}\label{sec:algorithm}
\name{} optimizes the \glspl{hris} based on an appropriately designed codebook (see Section \ref{sec:codebook}), which allows for the estimation of the \gls{bs}-\gls{ris} and \gls{ris}-\glspl{ue} \glspl{aoa} in a distributed manner. The use of codebooks is a known approach in \gls{ris}-assisted communications, e.g., \cite{WZ2020_TCOM,He2020}, and it is usually implemented by assuming that the electronic circuits of the \gls{ris} can realize a finite number of phase responses (e.g., through PIN diodes~\cite{boles2011algaas}). Therefore, our proposed \name{} is compatible with conventional implementations of \glspl{ris}, but it does not need a control channel.

Let us consider a codebook $\mathcal{C} = \{ \vv{c}_1,\dots,\vv{c}_\ssub{L}\}$, whose codewords $\vv{c}_l \in \mathbb{C}^{N\times 1}$ are unit-norm beamforming vectors that correspond to a discrete set of possible phase shift matrices $\vv{\Theta}_l = \mathrm{diag}(\vv{c}_l^{\herm})$. In particular, each codeword $\vv{c}_l$ is constituted by discrete-valued entries that mimic a sort of phase quantization. The discrete values of the codewords are assumed to belong to the following set 
\begin{equation}
    \mathcal{Q} = \bigg\{\frac{2\pi}{2^{Q}}m : m = 0,\dots, 2^{Q-1}, m \in \mathbb{N} \bigg\}, \label{eq:phase}
\end{equation}   
where $2^Q$ is the possible number of discrete values.

In \name{}, the \gls{hris} operates in two possible modes: \textit{probing} and \textit{communication}. In the probing mode, the \gls{hris} estimates the \glspl{aoa} that correspond to the \gls{bs} and to the \glspl{ue}. In the communication mode, the \gls{hris} assists the transmission of data between the \gls{bs} and the \glspl{ue}, while still being capable of detecting new \glspl{ue} joining the network or previously inactive \glspl{ue}. Therefore, we relax the assumption of simultaneous \glspl{ue} transmissions introduced in Section~\ref{sec:hris_config}, i.e., the \glspl{ue} may transmit at different times. Probing and communication phases are detailed in the following.

\begin{figure}[t]
        \center
        \includegraphics[width=\linewidth]{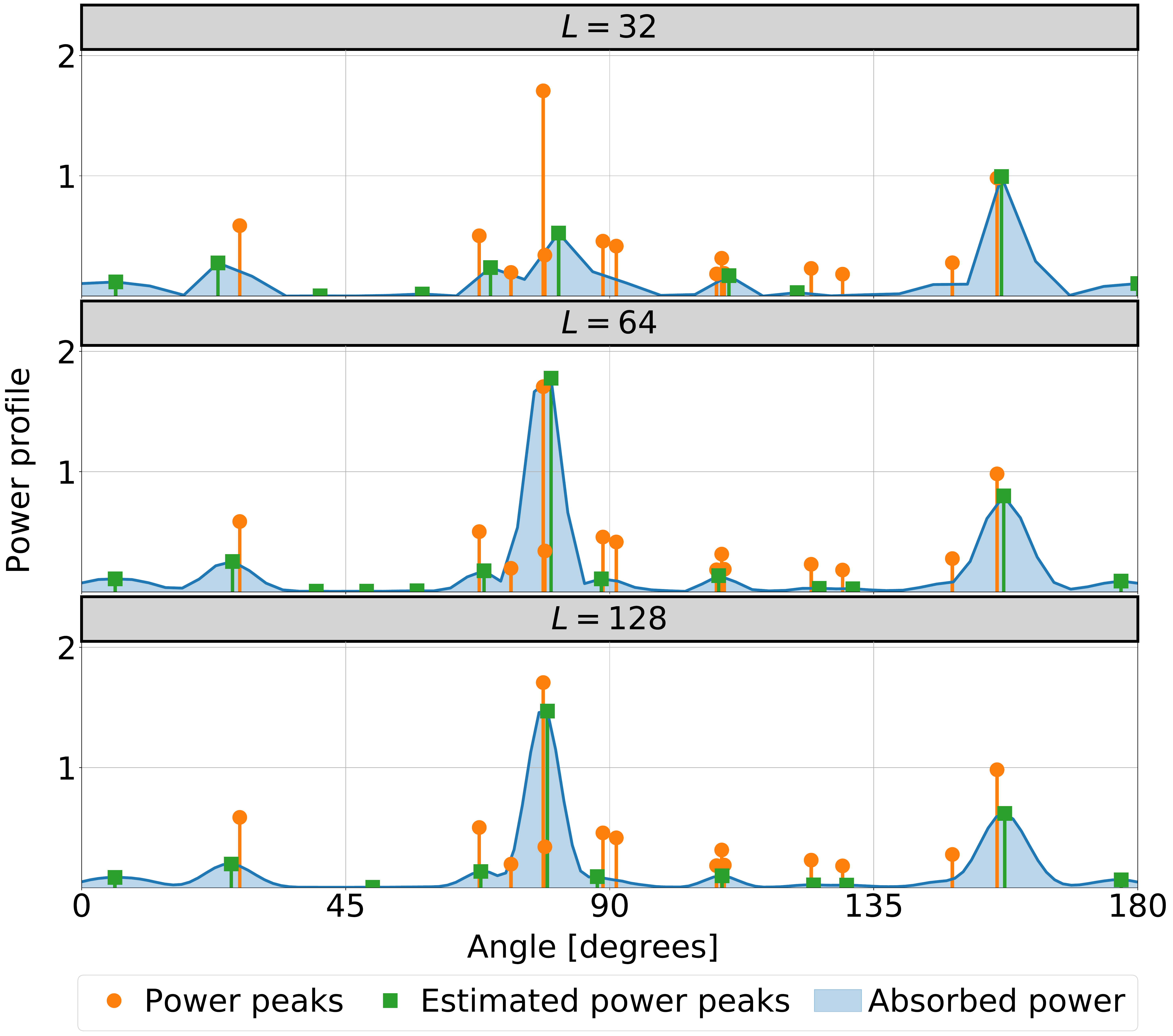}
        \caption{\label{fig:probing_phase} Example of power profile and corresponding estimated peak for different codebook sizes $L$ in a multi-\gls{ue} scenario with $K = 14$ \glspl{ue} and $N=32$ \gls{hris} elements.}
\end{figure}

\textbf{Probing phase.} Without loss of generality, we assume that each codeword of the codebook is, to a certain extent, spatially directive, i.e., the resulting \gls{hris} configuration maximizes the absorbed power only in correspondence of a (narrow) solid angle. This is relatively simple to realize by enforcing, e.g., some constraints on the design of the condewords in terms of half-power beamwidth of the corresponding radiation pattern of the \gls{hris}.
Therefore, by iteratively sweeping across all the codewords $\vv{c}_l \in \mathcal{C}$, the \gls{hris} can scan, with a given spatial resolution, the \gls{3d} space and can detect network devices (the \gls{bs} and the \glspl{ue}) by using pilot signals emitted only by those devices. During this probing phase, the \gls{hris} collects a set of power measurements, or equivalently a power profile, $\mathcal{P} = \{\rho_1,\dots,\rho_L\}$ where each element $\rho_l \in \mathbb{R}$ is the power level sensed (measured) by the \gls{hris} when using the codeword $\vv{c}_l$. As a result, the array response vectors in $\vv{G}$ or $\vv{h}_{\Sigma}$ can be estimated from $\mathcal{P}$. In practice, this boils down to detecting the peaks in $\mathcal{P}$ and identifying the corresponding angular directions. By construction, in fact, the \gls{hris} detects a power peak only if there is at least one transmitter in the direction synthesized by the  \gls{hris} beampattern (i.e., the considered codeword). The finer the angular selectivity of the \gls{hris}, the longer the probing phase. Therefore a suitable compromise needs to be considered. An example of power profile as a function of the steering angle of the \gls{hris} is reported in Fig.~\ref{fig:probing_phase}.

In particular, we assume that $\rho_l$ is a power peak in $\mathcal{P}$ if it is greater than a given threshold $\tau \in \mathbb{R}^+$. Let $\mathcal{I} \triangleq \{i < L \ : \ \rho_i \in \mathcal{P} > \tau \}$ be the set of indexes $l$ corresponding to the power peaks. 
Then, depending on which devices transmit their pilot signals, $\vv{v}_\ssub{B}$ and $\vv{v}_{\ssub{U}}$ in \eqref{eq:v_BU} can be estimated as
\begin{equation}
    \vv{v}_\ssub{B} = \sum\nolimits_{i \in \mathcal{I}} \delta_i \vv{c}_i, \quad \quad \vv{v}_{\ssub{U}} = \sum\nolimits_{i \in \mathcal{I}} \delta_i \vv{c}_i,
    \label{eq:v_B_code}
\end{equation} 
where $\delta_i \in \{1, \rho_i \}$ is a weight parameter that allows performing a hard ($\delta_i=1$) or a soft ($\delta_i=\rho_i$) combining of the power peaks in $\mathcal{I}$ based on the actual measured power $\rho_i$. 

The end-to-end \gls{hris} optimal configuration $\vv{v}_{\ssub{BU}}$ is first computed from \eqref{eq:v_BU} and is then projected onto the feasible set of discrete phase shifts in \eqref{eq:phase}, which eventually yields the desired  $\bar{\vv{v}}_{\ssub{BU}}$. The proposed probing phase is summarized in Algorithm~\ref{alg:marisa_prob}.

\textbf{Communication phase.} Upon completion of the probing phase, the \gls{hris} enters into the communication phase, which is aimed to assist the reliable transmission of data between the \gls{bs} and the active \glspl{ue}, as well as to probe the \gls{3d} space in order to discover new \glspl{ue}. The communication phase is summarized in Algorithm~\ref{alg:marisa_comm}. Once $\bar{\vv{v}}_{\ssub{BU}}$ is obtained, the \gls{hris} can easily aid the \gls{bs} and the active \glspl{ue} to communicate with each other. More challenging is the execution of the probing phase simultaneously with the communication phase. Therefore, hereafter we focus our attention on it.

To this end, we introduce the set of indices $\hat{\mathcal{I}} = \{l\in \mathbb{N}^+ : l\leq L\} \setminus \mathcal{I}$ that do not correspond to any power peaks estimated during the probing phase. Then, we construct the following codebook for simultaneous probing and communication
\begin{equation}
    \hat{\mathcal{C}} = \big\{ \mathrm{diag}(\bar{\vv{v}}_{\ssub{BU}}^{\herm}) + \vv{c}_j : \vv{c}_j \in \mathcal{C}, j \in \hat{\mathcal{I}}\big\},
\end{equation} whose codewords, after normalization and phase quantization, allow the \gls{hris} to scan the \gls{3d} space, looking for new \glspl{ue}, while keeping unaffected the beamsteering that corresponds to the \glspl{ue} being already served.

It is worth mentioning that the probing phase for discovering new \glspl{ue} does not replace the probing phase, executed on a regular basis, in Algorithm~\ref{alg:marisa_prob}. This is because it is necessary to check whether the estimated directions of the \gls{hris} still point towards the active \glspl{ue}.  

\begin{algorithm}[t!]
  \caption{\name{} -- Probing phase}\label{alg:marisa_prob}
  \begin{algorithmic}[1]
     \State Data: $\mathcal{C}$, $\tau \in \mathbb{R}^+$ 
     \State Perform a beam sweeping setting $\vv{\Theta}_l = \mathrm{diag}(\vv{c}_l)$, $\forall 
     \vv{c}_l \in \mathcal{C}$
     \State Measure the corresponding power profile $\mathcal{P}$
    \State Obtain $\mathcal{I} = \{i < L \ : \ \rho_i \in \mathcal{P} > \tau \}$
    \If{the \gls{bs} transmits the pilot signals}
    \State Compute $\vv{v}_{\ssub{B}} = \sum_{i \in \mathcal{I}} \delta_i \vv{c}_i^{\herm}$
    \ElsIf{the \glspl{ue} transmit the pilot signals}
    \State Compute    $\vv{v}_{\ssub{U}} = \sum_{i \in \mathcal{I}} \delta_i \vv{c}_i^{\herm}$
    \EndIf
    \State Obtain $\vv{v}_{\ssub{BU}} = \vv{v}_{\ssub{B}} \circ \vv{v}_{\ssub{U}}^*$ and $\bar{\vv{v}}_{\ssub{BU}}$ after quantization
    \end{algorithmic}
\end{algorithm}

\subsection{Codebook design}
\label{sec:codebook}
The probing and communication phases necessitate highly directive codebooks in order to scan the \gls{3d} space with a fine spatial resolution~\cite{devoti2020pasid,albanese2021papir}. To this end, we partition the total scanning angle  into $L$ angular sectors $[\phi_l,\phi_{l+1}]$ with $\phi_{l+1}-\phi_l = \pi/L$ $\forall l$. Each codeword $\vv{c}_l \in \mathcal{C}$ is, in particular, obtained as the solution of the following optimization problem  
\begin{problem}[Codebook design]\label{problem:codebook}
\begin{align}
   \max_{\vv{v}} &  \min_{\phi \in [\phi_l,\phi_{l+1}]} \  \|\vv{v}^{\herm} \vv{a}_{\ssub{R}}(\phi)\|^2 \nonumber\\
    \textup{s.t.} & \ \ \|\vv{v}^{\herm} \vv{a}_{\ssub{R}}(\phi)\|^2 \leq \epsilon , \quad \phi \in \{[0,\pi]\setminus[\phi_l, \phi_{l+1}]\} \nonumber \\
    & \ \ |v_i|^2 \leq  1 \quad  i=1,\dots,N, \nonumber 
\end{align}
\end{problem}
where, with a slight abuse of notation, $\vv{a}_{\ssub{R}}(\phi)$ denotes the \gls{hris} array response vector that points towards a \gls{ue} at an angle $\phi$ with respect to the \gls{hris} and that lies on the surface of a unit-radius sphere subtending the angle $\phi_{l+1}-\phi_l$, and $\epsilon \in \mathbb{R}^+$ denotes a threshold value to be finely tuned at the design stage (as further discussed in Section \ref{sec:results}). 


For fairness among the \glspl{ue}, we consider that the radiation pattern of the \gls{hris} across the generic $l$th angular sector is as flat as possible in $[\phi_l, \phi_{l+1}]$. 
Also, to minimize undesired reflections outside the angular range of interest, we enforce that the \gls{snr} of the \glspl{ue} in the complementary set $\{[0,\pi] \setminus [\phi_l, \phi_{l+1}]\}$ is below the threshold $\epsilon$.

By defining $\vv{V}\triangleq \vv{v}\vv{v}^{\herm}$, and $\overline{\vv{A}}(\phi)\triangleq\vv{a}_{\ssub{R}}(\phi)\vv{a}_{\ssub{R}}(\phi)^{\herm}$, Problem~\ref{problem:codebook} can be reformulated as follows.
\begin{problem}[Codebook design using \acrshort{sdr}]\label{problem:codebook_sdr}
\begin{align}
   \max_{\vv{V}\succeq\vv{0}} & \min_{\phi \in [\phi_l,\phi_{l+1}]} \tr(\vv{\overline{A}}(\phi)\vv{V}) \nonumber\\
   \textup{s.t.} & \ \ \tr(\vv{\overline{A}}(\phi)\vv{V}) \leq \epsilon^2, \quad \phi \in \{[0,\pi]\setminus[\phi_l, \phi_{l+1}]\} \nonumber\\ 
   & \ \ \mathrm{diag}(\vv{V}) \leq 1 \nonumber\\
   & \ \ \mathrm{rank}(\vv{V})=1. \nonumber
\end{align}
\end{problem}

Problem \ref{problem:codebook_sdr} can be relaxed by ignoring the non-convex rank constraint, and by employing the \gls{sdr} method in CVX~\cite{luo2010semidefinite, cvx}. Once the optimal solution  $\vv{V}_l$ of the relaxed version of Problem~\ref{problem:codebook_sdr} is obtained, a suboptimal solution to Problem~\ref{problem:codebook} is retrieved by using Gaussian randomization, which returns the desired codewords $\vv{c}_l$ after projection onto the feasible set $\mathcal{Q}$ in \eqref{eq:phase}.


\begin{algorithm}[t!]
  \caption{\name{} -- Communication phase}\label{alg:marisa_comm}
  \begin{algorithmic}[1]
     \State Data: $\mathcal{C}$, $\tau \in \mathbb{R}^+$ 
     \State Obtain $\hat{\mathcal{I}} = \{l\in \mathbb{N}^+ : l\leq L\} \setminus \mathcal{I}$
     \State Define $\hat{\mathcal{C}} = \big\{ \mathrm{diag}(\bar{\vv{v}}_{\ssub{BU}}^{\herm}) + \vv{c}_j : \vv{c}_j \in \mathcal{C}, j \in \hat{\mathcal{I}}\big\}$,
     \State Execute the probing phase in Algorithm~\ref{alg:marisa_prob} 
    \end{algorithmic}
\end{algorithm}

\section{Performance Evaluation}
\label{sec:results}
To prove the feasibility of \name{}, we evaluate it in different scenarios and compare it against the \gls{soa} benchmark scheme, recently reported in~\cite{Mursia2021}, which relies upon a control channel to perform a centralized optimization. 
The simulation setup and the parameters are given in Table~\ref{tab:parameters}. All results are averaged over $100$ simulation instances. 

\begin{table}[h!]
\caption{Simulation setup and parameters}
\label{tab:parameters}
\centering
\resizebox{\linewidth}{!}{%
\begin{tabular}{cc|cc|cc}
\textbf{Parameter} & \textbf{Value} & \textbf{Parameter} & \textbf{Value} & \textbf{Parameter} & \textbf{Value}\\  
\hline
\rowcolor[HTML]{EFEFEF}
$M$      & $4$             & $N_x,N_z$ & $8,4$     & $f_c$ & $28$                      GHz\\
$\vv{b}$ & $(-25,25,6)$ m  & $\vv{r}$  & $(0,0,6)$ m & $A$   & $50 \times 50$ $\text{m}^2$  \\
\rowcolor[HTML]{EFEFEF}
 $P$     & $20$ dBm        & $\beta_{\text{LoS}},\beta_{\text{NLoS}}$  & $2$, $4$  & $\sigma^2_n$  & $-80$ dBm \\
 $d_0,\gamma_0$   & $1$             & $\eta$      & $0.8$        &  $L$ & 32\\ 
\rowcolor[HTML]{EFEFEF}
 $\lambda_{\ssub{B}}$     & $0.3$ m$^{-2}$        & $h_{\ssub{B}}$  & $1.8$ m & $r_{\ssub{B}}$  & $0.6$ m \\
\end{tabular}%
}
\end{table}



The network area $A$ is a square, and the \gls{bs} and the \gls{hris} (or the \gls{ris}) are located in the midpoints of two of its adjacent edges. The \glspl{ue} are uniformly distributed in the network area, i.e., $\vv{u}_k \sim \mathcal{U}(A)$. To show the robustness of MARISA in realistic propagation scenarios, we relax the assumption of \gls{los} propagation conditions and account for the \gls{nlos} paths as well. In particular, we consider the stochastic geometry based model in~\cite{gapeyenko2017temporal}, which relates the geometric properties of the communication path in terms of the path length $l$ and height of the communicating devices (i.e., $u_z$, $b_z$, and $r_z$) in the presence of physical obstacles that may obstruct the links, which are referred to as blockers. The blockers are modeled as cylinders of height $h_{\ssub{B}}$, diameter $r_{\ssub{B}}$, and are distributed according to a \gls{ppp} with intensity $\lambda_{\ssub{B}}$. Therefore, for each path in the network area, the probability of \gls{nlos} propagation is
\begin{align}
    p(l)=1-e^{-2\lambda_{\ssub{B}} r_{\ssub{B}} \left(\sqrt{l^2-(b_z-u_z)^2}\frac{h_{\ssub{B}}-u_z}{h_{\ssub{B}}-u_z} + r_{\ssub{B}}\right)}.
\end{align}

The pathloss exponent for the \gls{los} or \gls{nlos} paths are denoted by $\beta_{\text{LoS}}$ or $\beta_{\text{NLoS}}$, respectively.


As far as the optimization of the precoder at the \gls{bs} is concerned, we assume perfect \gls{csi} at the \gls{bs}. In particular, the configuration of the \gls{hris} is assumed to be fixed by \name{} when optimizing the \gls{bs} precoder.
Therefore, the system is equivalent to a \gls{miso} channel given by the sum of the direct and reflected paths between the \gls{bs} and each \gls{ue}, where the \gls{hris} is viewed as an additional fixed scatterer (whose optimization is obtained by using \name{}). For a fair comparison with the benchmark scheme in \cite{Mursia2021}, the \gls{bs} precoder is chosen as
\begin{align}
    \vv{W}=\sqrt{P}\frac{\left( \vv{H}\vv{H}^{\herm} + \mu \vv{I}_m\right)^{-1}\vv{H}}{\|\left( \vv{H}\vv{H}^{\herm} + \mu \vv{I}_m\right)^{-1}\vv{H}\|_F},
\end{align}
where $\mu = K\frac{\sigma^2_n}{P}$, and each column of $\vv{H} \in \mathbb{C}^{M \times K}$ is the equivalent end-to-end \gls{miso} channel between the \gls{bs} and the corresponding \gls{ue}. 

It is worth mentioning that the performance of centralized deployments and \name{} depend on the overhead for channel estimation and reporting \cite{zappone2020overhead}, and the overhead of the probing phase \cite{rouissi2015design}, respectively. These two solutions are very different from each other and a fair comparison of the associated overhead is postponed to a future research work.

\subsection{Comparison with the Centralized Deployment}

\begin{figure}[t]
        \center
        \includegraphics[width=\linewidth, trim = {0cm, 4cm, 0cm, 0cm}]{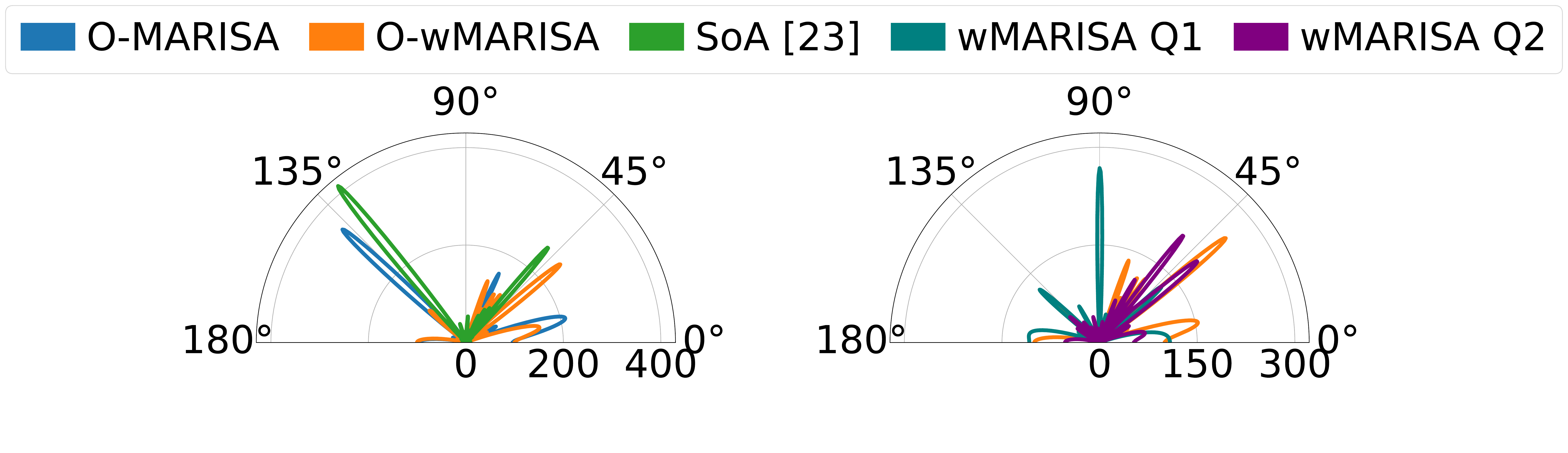}
        \caption{\label{fig:beam_pattern_marisa}
        Radiation pattern at the \gls{hris} along the azimuth directions obtained with O-\name{}, O-w\name{}, and \gls{soa} (left), and radiation patterns obtained with $Q$ bits of phase quantization (right).}
\end{figure}

\begin{figure}[t]
        \center
        \includegraphics[width=\linewidth]{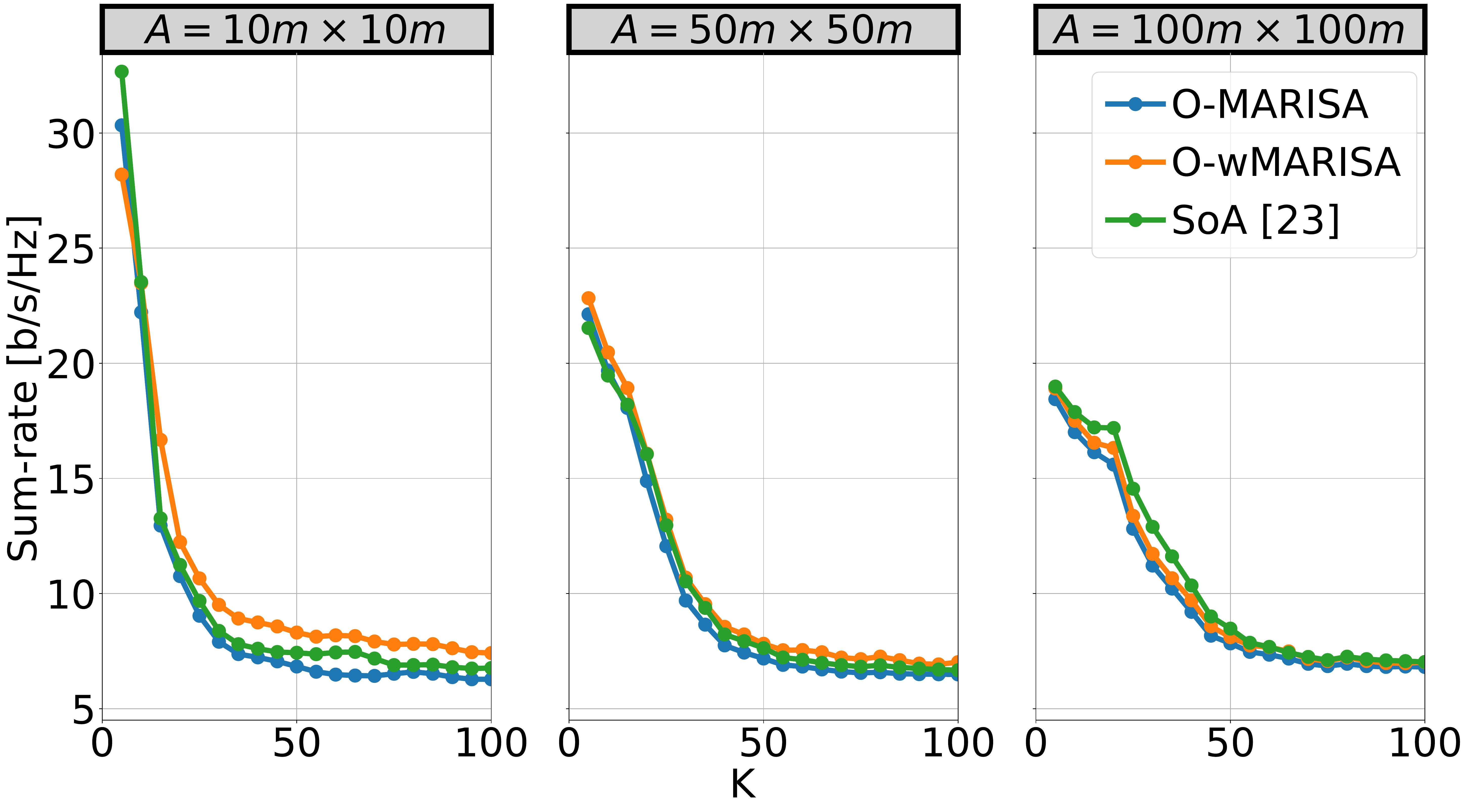}
        \vspace{-7mm}
        \caption{\label{fig:oracle_vs_SoA_K} Average sum-rate in a multi-\gls{ue} scenario obtained by solving Problem~\ref{problem:max_sinr_multi_ue} with perfect \gls{csi} and by \gls{soa}~\cite{Mursia2021} against the number of \glspl{ue} $K$ for different network areas and when the number of \gls{hris} elements is $N=32$.}
\end{figure}
\begin{figure}[t]
        \center
        \includegraphics[width=\linewidth]{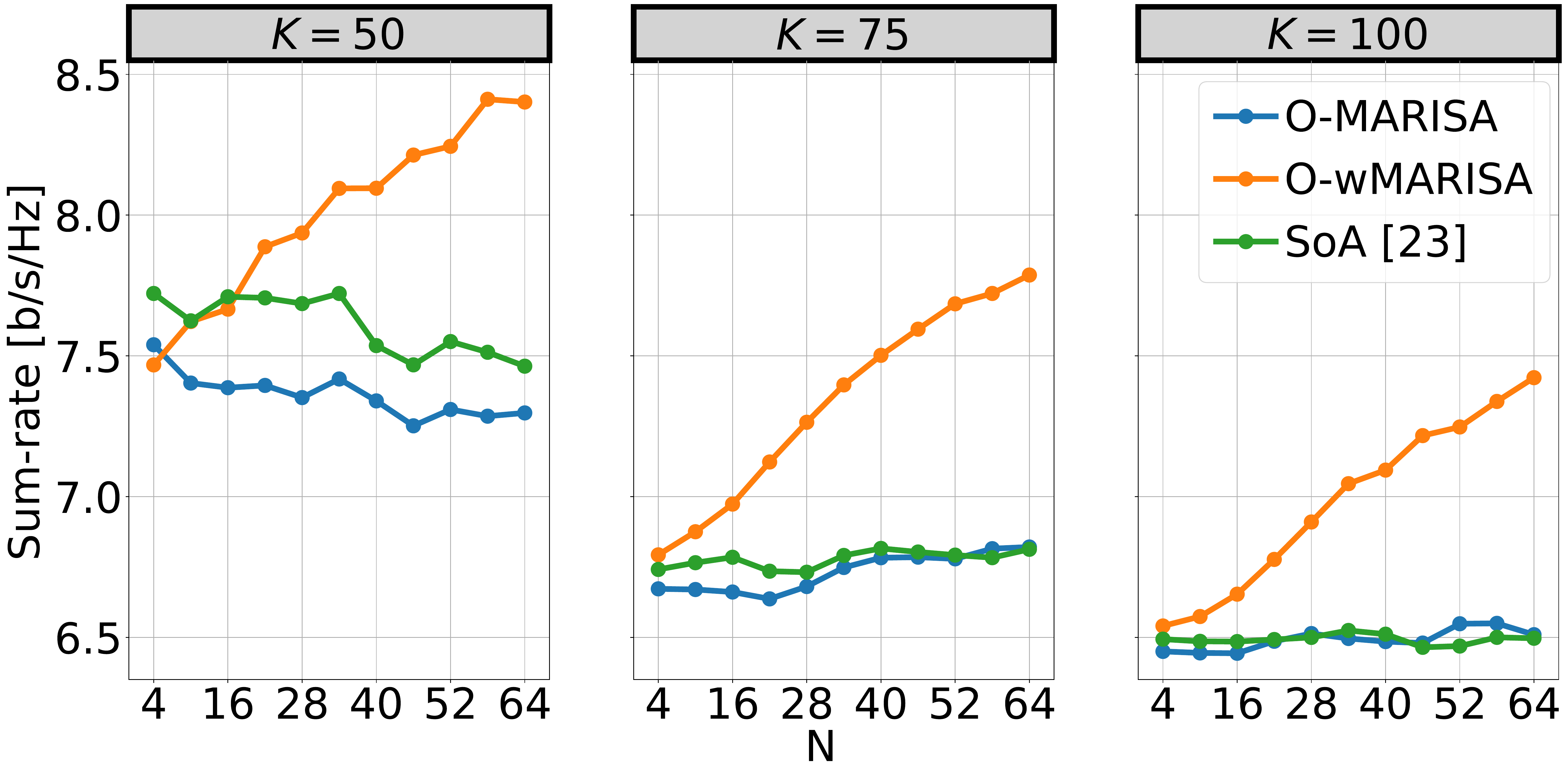}
        \vspace{-7mm}
        \caption{\label{fig:oracle_vs_SoA_N} Average sum-rate in a multi-\gls{ue} scenario obtained by solving Problem~\ref{problem:max_sinr_multi_ue} with perfect \gls{csi} and by \gls{soa}~\cite{Mursia2021} against the number of \gls{hris} elements $N$, the number of \glspl{ue} $K$. The network area is $A = 50$ m $\times$ $50$ m.}
\end{figure}

We analyze the viability of self-configuring an \gls{hris} by solving Problem~\ref{problem:max_sinr_multi_ue} with perfect knowledge of the aggregate \gls{hris}-\gls{ue} channel $\vv{h}_{\ssub{\Sigma}}$ and of the response vector of the \gls{hris} towards the \gls{bs} $\vv{a}_{\ssub{R}}(\vv{b})$ in~\eqref{eq:v_BU}. We refer to this design as the \textit{Oracle} (O) scheme, since the channels are known already and do not need to be estimated.

Moreover, we analyze two solutions that assume real-valued (continuous) phase shifts: $i$) O-\name{}, which calculates $\vv{h}_{\ssub{\Sigma}} = \sum_k \frac{\vv{h}_{k}}{\|\vv{h}_{k}\|}$, and $ii$) O-weighted \name{} (O-w\name{}), which calculates $\vv{h}_{\ssub{\Sigma}} = \sum_k \vv{h}_{k}$. Specifically, O-\name{} estimates $\vv {v}_{\ssub{B}}$ and $\vv{v}_{\ssub{U}}$ only based on the direction of the paths that are assumed to have a unit gain, while O-w\name{} utilizes the direction and the gain of the paths.

Figure~\ref{fig:beam_pattern_marisa} (left) shows a comparison of the \gls{hris} configuration obtained by O-\name{}, O-w\name{}, and the \gls{soa} centralized solution in~\cite{Mursia2021}, which jointly optimizes the \gls{bs} precoder and the \gls{ris} phase shifts by means of a control channel. 
While the \gls{soa} provides a very directive beampattern with few enhanced directions, both versions of O-\name{} result in a wider range of directions at the expense of a smaller gain due to the presence of multiple secondary lobes. Despite the different beampatterns, the sum-rates obtained by \name{} and the centralized benchmark, as shown in Fig.~\ref{fig:oracle_vs_SoA_K}, are very similar. In particular, O-\name{} and the \gls{soa} provide a sum-rate that does not increase with the number of \glspl{ue}, which hints to an interference-constrained scenario. Notably, O-w\name{} delivers better performance thanks to the weighting mechanism that strengthens the reflected paths with higher power gains. This behavior is further confirmed in Fig.~\ref{fig:oracle_vs_SoA_N}, where the average sum-rate is analyzed against the number of \gls{hris} elements $N$ and \glspl{ue} $K$. We see that, in this case, only O-w\name{} has a non-decreasing behavior regardless of the interference-constrained nature of the scenario, largely outperforming the \gls{soa} centralized solution in~\cite{Mursia2021}.



\subsection{Codebook-Based \name{}}

In this section, we analyze the performance offered by \name{} under the realistic assumption that the \gls{hris} optimizes its configuration through power measurements and by iteratively activating the beam patterns (codewords) in the codebook $\mathcal{C}$. Therefore, no apriori knowledge of the aggregate channel $\vv{h}_{\ssub{\Sigma}}$ and of the response vector towards the \gls{bs} $\vv{a}_{\ssub{R}}(\vv{b)}$ is assumed. Also, the phase shifts applied by the \gls{hris} belong to the discrete set $\mathcal{Q}$ in \eqref{eq:phase}. 
The steering directions are computed based on the estimated peaks in the measured power profile $\mathcal{P}$.
Supported by the previous case study, we analyze only the performance of w\name{}. Based on the estimated angular power profile  $\mathcal{P}$, $\vv{v}_{\ssub{B}}$ and $\vv{v}_{\ssub{U}}$ are estimated from \eqref{eq:v_B_code}. The weights $\delta_l$ are set equal to $\rho_l$,  $\forall l \in \mathcal{I}$ and $0$, $\forall l \in \hat{\mathcal{I}}$.


\begin{figure}[t]
        \center
        \includegraphics[width=\linewidth]{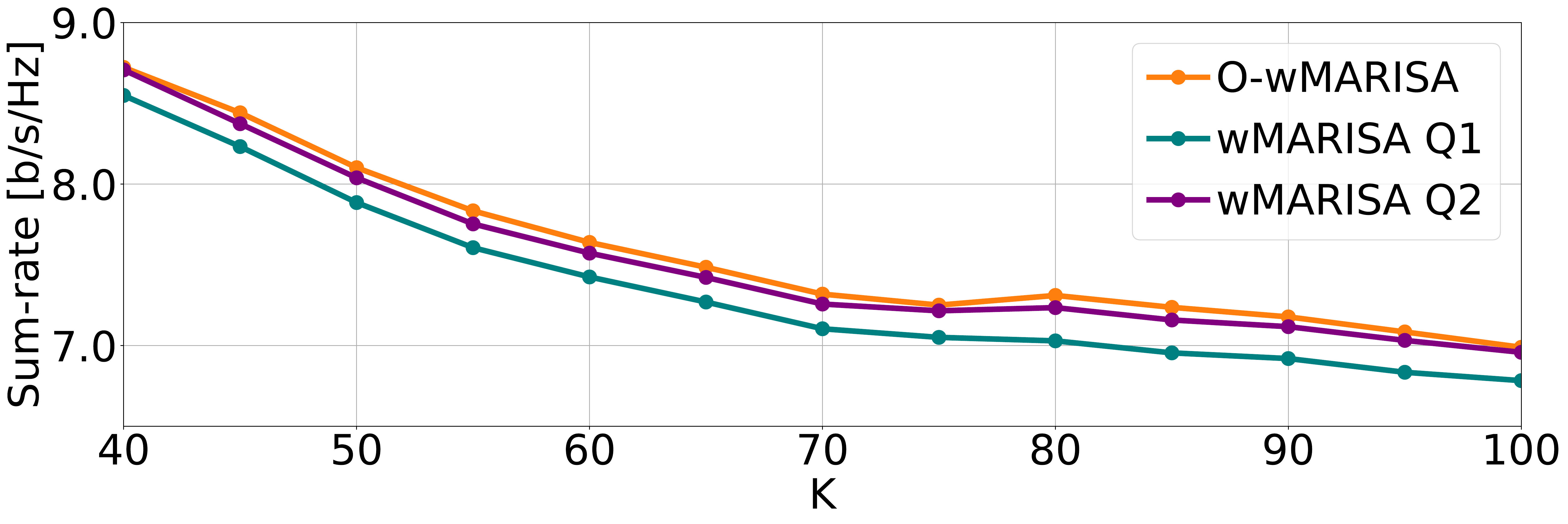}
        \vspace{-7mm}
        \caption{\label{fig:oracle_vs_est_K}
        Average sum-rate in a multi-\gls{ue} scenario obtained by w\name{} for different quantization levels $Q$, and O-w\name{}, against the number of \glspl{ue} $K$.}
        \vspace{-1.9mm}
\end{figure}
\begin{figure}[t]
        \center
        \includegraphics[width=\linewidth]{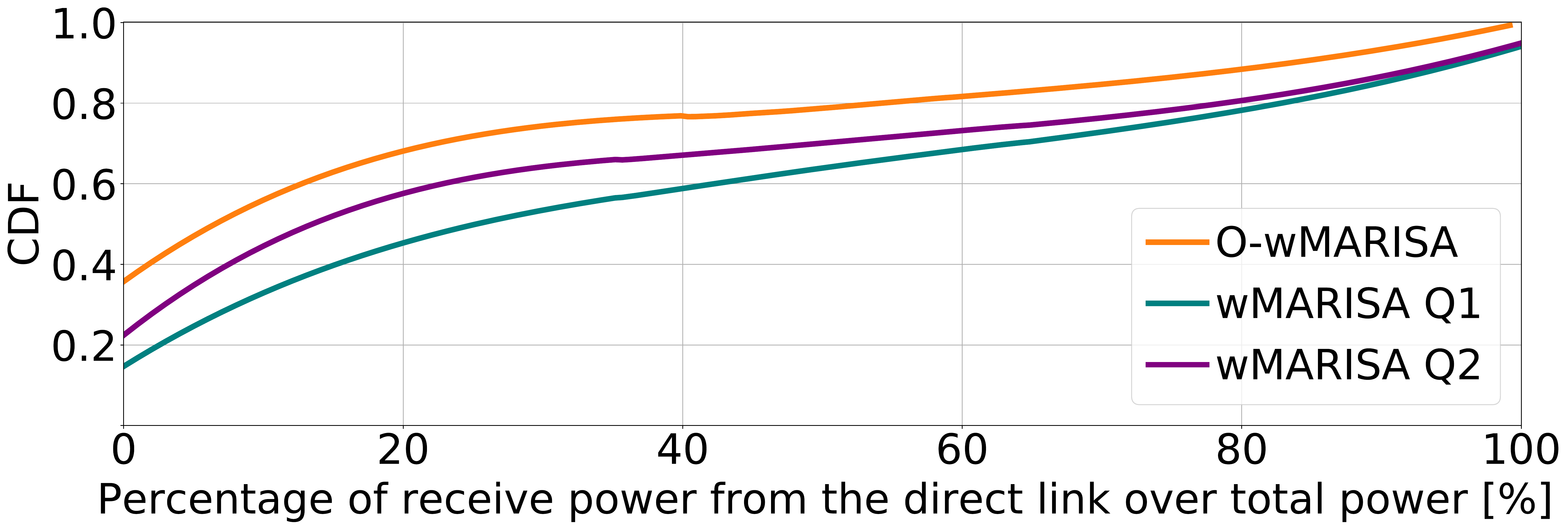}
        \vspace{-6.8mm}
        \caption{\label{fig:pdf} Cumulative distribution function (CDF) of the fraction of the receive power at each \gls{ue} over the direct path with respect to the total receive power after precoder optimization and selection at the \gls{bs}.}
\end{figure}

We consider two implementations for w\name{}, denoted by w\name{} Q1 and w\name{} Q2, which correspond to the w\name{} algorithm with the quantization levels $Q=1$ and $Q=2$ bits, respectively. The achievable average sum-rate is reported in Fig.~\ref{fig:oracle_vs_est_K}. Relaxing the assumptions of perfect \gls{csi} and continuous phase shifts has only a limited impact on the sum-rate, which confirms the effectiveness of the approach proposed in Section~\ref{sec:overview}. As expected, the sum-rate worsens when one quantization bit is used, while two quantization bits offer good performance already.

In Fig.~\ref{fig:pdf}, finally, we report the distribution of the percentage of power that every \gls{ue} receives from the direct link with respect to the total received power (from the direct link and the reflected link). We note that O-w\name{} offers the highest power boost that originates from the reflect paths thanks to its ideal beamforming capabilities. On the other hand,  w\name{} Q$1$ and Q$2$ are affected by quantization errors that lead to  beampatterns with a more distributed power spread. Similar unwanted reflections can be seen in Fig.~\ref{fig:beam_pattern_marisa} (right), where the beampatterns obtained with O-w\name{} and w\name{} Q$1$ and Q$2$ are reported. However, these unwanted reflections have little impact on the sum-rate.

\section{Related Works}
\label{sec:related}
In the last few years, \glspl{ris} have drawn considerable interest from the scientific community due to their ability to turn uncontrollable propagation channels into controlled variables that can be optimized~\cite{LZWXA21_LCOMM,LFY21_Access}. A preliminary analysis of the achievable performance of an \gls{ris} is given in~\cite{WZ2019_TWC}. In particular, the authors formulate a joint optimization problem for optimizing the active beamforming (at the multi-antenna \gls{bs}) and the passive beamforming configuration (at the \gls{ris}), and they demonstrate that \gls{ris}-based \gls{mimo} systems can achieve rate performance similar to legacy massive \gls{mimo} systems with fewer active antenna elements. The ideal case study with continuous phase shifts at the \gls{ris} is generalized to the case with discrete phase shifts in~\cite{WZ2020_TCOM}. The authors prove the squared power gain with the number of reflecting elements even in the presence of phase quantization, but a power loss that depends on the number of phase-shift levels is observed~\cite{yub21_TCOM}. In~\cite{Mursia2021}, the authors propose a practical algorithm to maximize the system sum mean squared error while jointly optimizing the transmit beamforming at \gls{mimo} \glspl{bs} and the \gls{ris} configuration. 

Other papers have recently considered the possibility of optimizing the \glspl{ris} based on statistical \gls{csi} in order to relax the associated feedback overhead. A two-timescale transmission protocol is considered in~\cite{zhao2020intelligent} to maximize the achievable average sum-rate for an \gls{ris}-aided multiuser system under a general correlated Rician channel model, whereas~\cite{ADD21_TCOM} and \cite{zprw21_LWC} maximize the network sum-rate by means of the statistical characterization of the locations of the users, which does not require frequent updates of the \gls{ris} reconfiguration. These solutions, however, rely on the presence of a control channel.

A detailed analysis of an \gls{ris}-assisted multi-stream \gls{mimo} system is described in~\cite{PTDF2021_TWC}, where the authors formulate a joint optimization problem of the covariance matrix of the transmitted signal and the \gls{ris} phase shifts. An effective solution is obtained, which offers similar performance to \gls{soa} schemes but with limited computational complexity. The approach is generalized in \cite{PTDF2021_LWC} to discrete-valued constellation symbols. A comprehensive tutorial on \glspl{ris} focused on optimization is available in~\cite{WZZYZ2021_TCOM}.

None of above-mentioned works deals with self-configuring \gls{ris}-empowered networks without relying on a control channel, which is, on the other hand, the main contribution and novelty of the present paper. 

\section{Conclusions}
\label{sec:conclusion}
\glspl{ris} are an emerging clean-slate technology with the inherent potential of fundamentally reshaping the design and deployment of mobile communication systems. In this paper, we introduced \name{}, a Metasurface Absorption and Reflection solution for Intelligent Surfaces Applications, which treats \glspl{hris} as plug-and-play devices that are endowed with joint reflection and power-sensing capabilities. \name{} is built upon $i$) a new channel estimation model lato-sensu at the \gls{ris} and $ii$) an autonomous \gls{ris} configuration methodology that is based only on \gls{csi} that can be locally estimated at the \glspl{hris}, without requiring an active control channel. Our results unveil promising performance trends: \name{} provides \emph{near-optimal sum-rates when compared to fully \gls{csi}-aware benchmark schemes that rely on a dedicated control channel}. 


\bibliographystyle{IEEEtran}
\bibliography{references}


\end{document}